\documentclass[twocolumn]{aastex62}
\usepackage{natbib,amsmath,graphicx,enumerate,epsfig}
\usepackage{color}

\def\ts {\thinspace}
\def\kms {\ifmmode{{\rm \ts km\ts s}^{-1}}\else{\ts km\ts s$^{-1}$}\fi}
\def\ergs {\ifmmode{{\rm \ts erg\ts s}^{-1}}\else{\ts erg\ts s$^{-1}$}\fi}
\def\ergscc {\ifmmode{{\rm \ts erg\ts s}^{-1}{\rm \ts cm}^{-2}{\rm \ts sr}^{-1}}\else{\ts erg\ts s$^{-1}$\ts cm$^{-2}$\ts sr$^{-1}$}\fi}
\def\ergsccas {\ifmmode{{\rm \ts erg\ts s}^{-1}{\rm \ts cm}^{-2}{\rm \ts as}^{-2}}\else{\ts erg\ts s$^{-1}$\ts cm$^{-2}$\ts as$^{-2}$}\fi}
\def\kmspc {\ifmmode{{\rm \ts km\ts s}^{-1}\ts{\rm pc}^{-1}}\else{\ts km\ts s$^{-1}$\ts pc$^{-1}$}\fi}
\def\kkms {\ifmmode{{\rm \ts K\ts km\ts s}^{-1}}\else{\ts K\ts km\ts s$^{-1}$}\fi}
\def\lcou {\ifmmode{{\rm \ts K\ts km\ts s}^{-1}\ts {\rm pc}^{2}}\else{\ts K\ts km\ts s$^{-1}$\ts pc$^{2}$}\fi}
\def\xco {\ifmmode{X_{\rm CO}}\else{$X_{\rm CO}$}\fi}
\def\xcou {\ifmmode{{\rm \ts cm^{-2}\ts (K\ts km\ts s^{-1})}^{-1}}\else{\ts cm$^{-2}$\ts (K\ts km\ts s$^{-1}$)$^{-1}$}\fi}
\newcommand{\acou}{\mbox{M$_\odot$ pc$^{-2}$ (K km s$^{-1}$)$^{-1}$}}
\def\phu {\ifmmode{{\rm \ts 10^{4}\ts k_{B}\ts cm^{-3}\ts K}}\else{\ts 10$^{4}$\ts k$_{B}$\ts cm$^{-3}$\ts K}\fi}
\def\cc {\ifmmode{{\rm \ts cm}^{-2}}\else{\ts cm$^{-2}$}\fi}
\def\ccc {\ifmmode{{\rm \ts cm}^{-3}}\else{\ts cm$^{-3}$}\fi}
\def\mo {\ifmmode{{\rm M}_{\odot}}\else{M$_{\odot}$}\fi}
\def\msol {\ifmmode{{\rm M}_{\odot}}\else{M$_{\odot}$}\fi}
\def\lsol {\ifmmode{{\rm L}_{\odot}}\else{L$_{\odot}$}\fi}
\def\mpcsq {\ifmmode{{\rm M}_{\odot}\ts {\rm pc}^{-2}}\else{M$_{\odot}$}\ts pc$^{-2}$\fi}
\def\myrkpcsq {\ifmmode{{\rm M}_{\odot}\ts {\rm yr}^{-1}\ts {\rm kpc}^{-2}}\else{M$_{\odot}$}\ts \ts yr$^{-1}$\ts kpc$^{-2}$\fi}
\def\myr {\ifmmode{{\rm M}_{\odot}\ts {\rm yr}^{-1}}\else{M$_{\odot}$}\ts yr$^{-1}$\fi}
\def\msolyr {\ifmmode{{\rm M}_{\odot}\ts {\rm yr}^{-1}}\else{M$_{\odot}$}\ts yr$^{-1}$\fi}

\def\rco {\ifmmode{R_{21}}\else{$R_{21}$}\fi}
\def\aco {\ifmmode{^{12}{\rm CO}(J=1\to0)}\else{$^{12}{\rm CO}(J=1\to0)$}\fi}
\def\bco {\ifmmode{^{12}{\rm CO}(J=2\to1)}\else{$^{12}{\rm CO}(J=2\to1)$}\fi}
\def\cco {\ifmmode{^{13}{\rm CO}(J=1\to0)}\else{$^{13}{\rm CO}(J=1\to0)$}\fi}
\def\dco {\ifmmode{^{13}{\rm CO}(J=2\to1)}\else{$^{13}{\rm CO}(J=2\to1)$}\fi}
\def\eco {\ifmmode{{\rm C}^{18}{\rm O}(J=1\to0)}\else{{\rm C}$^{18}{\rm O}(J=1\to0)$}\fi}
\def\mu {\ifmmode{\mu {\rm m}}\else{$\mu$m}\fi}
\def\hi  {\ifmmode{{\rm H}{\rm \scriptsize I}}\else{H\ts {\scriptsize I}}\fi}
\def\nii  {\ifmmode{{\rm N}{\rm \scriptsize II}}\else{N\ts {\scriptsize II}}\fi}
\def\sii  {\ifmmode{{\rm S}{\rm \scriptsize II}}\else{S\ts {\scriptsize II}}\fi}
\def\hii  {\ifmmode{{\rm H}{\rm \scriptsize II}}\else{H\ts {\scriptsize II}}\fi}
\def\hh  {\ifmmode{\rm H_{\rm 2}}\else{H$_{\rm 2}$}\fi}
\def\Ha  {\ifmmode{{\rm H}{\alpha}}\else{H\ts {$\alpha$}}\fi}
\def\Halpha  {\ifmmode{{\rm H}{\alpha}}\else{H\ts {$\alpha$}}\fi}
\def\ha  {\ifmmode{{\rm H}{\alpha}}\else{H{$\alpha$}}\fi}
\def\halpha  {\ifmmode{{\rm H}{\alpha}}\else{H\ts {$\alpha$}}\fi}
\def\Hb  {\ifmmode{{\rm H}{\beta}}\else{H\ts {$\beta$}}\fi}
\def\Hbeta  {\ifmmode{{\rm H}{\beta}}\else{H\ts {$\beta$}}\fi}
\def\hb  {\ifmmode{{\rm H}{\beta}}\else{H\ts {$\beta$}}\fi}
\def\hbeta  {\ifmmode{{\rm H}{\beta}}\else{H\ts {$\beta$}}\fi}
\def\nh  {\ifmmode{N(\rm H)}\else{$N$(H)}\fi}
\def\nhi  {\ifmmode{N(\hi)}\else{$N$(\hi)}\fi}
\def\nhh  {\ifmmode{N(\rm H_{\rm 2})}\else{$N$(H$_{\rm 2}$)}\fi}
\def\nhm  {\ifmmode{N({\rm H})_{\rm mol}}\else{$N$(H)$_{\rm mol}$}\fi}
\def\hh   {\ifmmode{{\rm H}_2}\else{H$_2$}\fi}
\def\ico {\ifmmode{I(\rm CO)}\else{$I(\rm CO)$}\fi}
\def\lco {\ifmmode{L_{\rm CO}}\else{$L_{\rm CO}$}\fi}
\def\tpk {\ifmmode{T_{\rm pk}}\else{$T_{\rm pk}$}\fi}
\def\tpkco {\ifmmode{T_{\rm pk}(\rm CO)}\else{$T_{\rm pk}(\rm CO)$}\fi}
\def\D {\ifmmode{^{\circ}}\else{$^\circ$}\fi}
\def\sings {\ifmmode{{\rm SINGS}}\else{SINGS}\fi}
\def\sfr {\ifmmode{{\rm SFR}}\else{SFR}\fi}
\def\alphaco {\ifmmode{\alpha_{\rm CO}}\else{$\alpha_{\rm CO}$}\fi}
\def\acou {\ifmmode{\ts \mpcsq \ts (K\ts km\ts s^{-1})}^{-1}\else{\ts \mpcsq \ts (K\ts km\ts s$^{-1}$)$^{-1}$}\fi}
\def\mic {\ifmmode{\mu {\rm m}}\else{$\mu$m}\fi}

\def\comm#1 {{\tt (COMMENT: #1) }}

\shorttitle{Gas and Star Formation in PHANGS Galaxies}
\shortauthors{Schinnerer, Hughes, Leroy, Groves \& PHANGS}

\begin{document}

\title{The Gas-Star Formation Cycle in Nearby Star-Forming Galaxies I. Assessment of Multi-scale Variations}

\author{Eva Schinnerer}
\affiliation{Max Planck Institute for Astronomy, K\"{o}nigstuhl 17,
\\
 69117 Heidelberg, Germany}
\affiliation{Adjunct Scientist, National Radio Astronomy Observatory, 
\\
520 Edgemont Road, Charlottesville, VA 22903, U.S.A.}

\author{Annie Hughes}
\affiliation{Universit\'{e} de Toulouse, UPS-OMP, 31028 Toulouse, France}
\affiliation{CNRS, IRAP, Av. du Colonel Roche BP 44346, 31028 Toulouse 
\\
cedex 4, France}

\author{Adam Leroy}
\affiliation{Department of Astronomy, The Ohio State University, 
\\
140 West 18th Ave, Columbus, OH 43210, USA}

\author{Brent Groves}
\affiliation{Research School of Astronomy and Astrophysics, Australian
\\
National University, Canberra, ACT 2611, Australia}
\affiliation{International Centre for Radio Astronomy Research, The 
\\
University of Western Australia, Crawley, WA 6009 Australia}

\author{Guillermo A. Blanc}
\affiliation{The Observatories of the Carnegie Institution for Science,
\\
813 Santa Barbara Street, Pasadena, CA 91101, USA} 
\affiliation{Departamento de Astronom\'{i}a, Universidad de Chile,
\\
Casilla 36-D, Santiago, Chile}

\author{Kathryn Kreckel}
\affiliation{Max Planck Institute for Astronomy, K\"{o}nigstuhl 17,
\\
69117 Heidelberg, Germany}

\author{Frank Bigiel}
\affiliation{Argelander-Institut f\"{u}r Astronomie, Universit\"{a}t Bonn, 
\\
Auf dem H\"{u}gel 71, 53121 Bonn, Germany}

\author{M\'{e}lanie Chevance}
\affiliation{Astronomisches Rechen-Institut, Zentrum f\"{u}r Astronomie der Universit\"{a}t
\\
Heidelberg, M\"{o}nchhofstra\ss e 12-14, 69120 Heidelberg, Germany}

\author{Daniel Dale}
\affiliation{Department of Physics and Astronomy, University of Wyoming,
\\
Laramie, WY 82071, USA}

\author{Eric Emsellem}
\affiliation{European Southern Observatory, Karl-Schwarzschild-Strasse 2,
\\
D-85748 Garching bei M\"{u}nchen, Germany}
\affiliation{Univ. Lyon, Univ. Lyon1, ENS de Lyon, CNRS, Centre de Recherche 
\\
Astrophysique de Lyon UMR5574, F-69230 Saint-Genis-Laval, France}

\author{Christopher Faesi}
\affiliation{Max Planck Institute for Astronomy, K\"{o}nigstuhl 17,
\\
69117 Heidelberg, Germany}

\author{Simon Glover}
\affiliation{Instit\"ut  f\"{u}r Theoretische Astrophysik, Zentrum f\"{u}r Astronomie der Universit\"{a}t
\\
Heidelberg, Albert-Ueberle-Strasse 2, 69120 Heidelberg, Germany}

\author{Kathryn Grasha}
\affiliation{Research School of Astronomy and Astrophysics, Australian
\\
National University, Canberra, ACT 2611, Australia}

\author{Jonathan Henshaw}
\affiliation{Max Planck Institute for Astronomy, K\"{o}nigstuhl 17,
\\
69117 Heidelberg, Germany}

\author{Alexander Hygate}
\affiliation{Astronomisches Rechen-Institut, Zentrum f\"{u}r Astronomie der Universit\"{a}t 
\\
Heidelberg, M\"{o}nchhofstra\ss e 12-14, 69120 Heidelberg, Germany}

\author{J.~M.~Diederik Kruijssen}
\affiliation{Astronomisches Rechen-Institut, Zentrum f\"{u}r Astronomie der Universit\"{a}t
\\
Heidelberg, M\"{o}nchhofstra\ss e 12-14, 69120 Heidelberg, Germany}

\author{Sharon Meidt}
\affiliation{Sterrenkundig Observatorium, Universiteit Gent, Krijgslaan 281 S9, B-9000 Gent, Belgium}

\author{Jerome Pety}
\affiliation{IRAM, 300 rue de la Piscine, F-38406 Saint Martin de H\`{e}res, France}
\affiliation{Sorbonne Universit\'{e}, Observatoire de Paris, Universit\'{e} PSL, \'{E}cole normale sup\'{e}rieure,
\\
CNRS, LERMA, F-75005, Paris, France}

\author{Miguel Querejeta}
\affiliation{European Southern Observatory, Karl-Schwarzschild-Strasse 2,
\\
D-85748 Garching bei M\"{u}nchen, Germany}
\affiliation{Observatorio Astron{\'o}mico Nacional (IGN),
\\
C/Alfonso XII 3, Madrid E-28014, Spain}

\author{Erik Rosolowsky}
\affiliation{4-183 CCIS, University of Alberta, Edmonton, Alberta, Canada}

\author{Toshiki Saito}
\affiliation{Max Planck Institute for Astronomy, K\"{o}nigstuhl 17,
\\
69117 Heidelberg, Germany}

\author{Andreas Schruba}
\affiliation{Max-Planck-Institut f\"{u}r extraterrestrische Physik,
\\
Giessenbachstra\ss e 1, D-85748 Garching, Germany}

\author{Jiayi Sun}
\affiliation{Department of Astronomy, The Ohio State University, 
\\
140 West 18th Ave, Columbus, OH 43210, USA}

\author{Dyas Utomo}
\affiliation{Department of Astronomy, The Ohio State University,
\\
140 West 18th Ave, Columbus, OH 43210, USA}

\begin{abstract}
The processes regulating star formation in galaxies are thought to act
across a hierarchy of spatial scales. To connect extragalactic star
formation relations from global and kpc-scale measurements to recent
cloud-scale resolution studies, we have developed a simple, robust
method that quantifies the scale dependence of the relative spatial
distributions of molecular gas and recent star formation.  In this
paper, we apply this method to eight galaxies with $\sim1\arcsec$
resolution molecular gas imaging from the PHANGS-ALMA and PAWS surveys
that have matched resolution, high quality narrowband H$\alpha$
imaging. At a common scale of 140\,pc, our massive ($\rm
log(M_{\star}[M_{\odot}])\,=\,9.3-10.7$), normally star-forming ($\rm
SFR[M_{\odot}\,yr^{-1}]\,=\,0.3-5.9$) galaxies exhibit a significant
reservoir of quiescent molecular gas not associated with star
formation as traced by \ha\ emission. Galactic structures act as backbones for both 
molecular and \hii\ region distributions. As we degrade the spatial resolution, the
quiescent molecular gas disappears, 
with the most rapid changes occurring for resolutions up to
$\sim$\,0.5\,kpc. As the resolution becomes poorer, the morphological features become
indistinct for spatial scales larger than $\sim1$\,kpc. The method is a
promising tool to search for relationships between the quiescent or
star-forming molecular reservoir and galaxy properties, but requires a
larger sample size to identify robust correlations between the
star-forming molecular gas fraction and global galaxy parameters.
\end{abstract}

  

\section{Introduction}

The star formation rate of galaxies reflects the interplay of galactic
dynamics, violent stellar feedback (also from active galactic nuclei
(AGN) if present), and gravitational collapse of molecular
gas. Accessing these physics requires high resolution observations
spanning large areas across a diverse sample of galaxies.

Observations at coarse resolution, from $\sim 1$~kpc scales up to
whole galaxies, show a tight correlation between molecular gas and
star formation \citep[e.g.,][among many
  others]{young95,kennicutt98b,wong02,leroy08,bigiel08,schruba11,saintonge11,leroy13,momose13}. To the extent that galactic morphology remains discernible at these scales, such studies have typically found that maps of molecular gas and star formation tracers are very similar.

High resolution observations suggest a more complex picture. When the
resolution approaches the scale of individual giant molecular clouds
(GMCs) and {\sc Hii} regions, tracers of recent massive star formation
and cold molecular gas appear spatially distinct
\citep[e.g.,][]{kawamura09,schruba10,gratier12,battersby17,kreckel18,querejeta19}. This
separation between the input (GMCs) and output ({\sc Hii} regions) of the
star formation process has been interpreted as a sign of destructive feedback
and used to estimate molecular cloud lifetimes
\citep[e.g.,][]{kawamura09,schruba10,gratier12,kruijssen14,battersby17,kruijssen18,kruijssen19}. A related measurement shows that the scatter in star formation--gas
scaling relations increases with improving spatial resolution
\citep[e.g.,][]{bigiel08,blanc09,schruba10,onodera10,leroy13,kreckel18}. This,
too, has been interpreted as a signature of evolutionary cycling
and/or feedback
\citep[e.g.,][]{feldmann11,kruijssen14,kruijssen18}. Observations at high spatial resolution can potentially measure the impact of stellar feedback
on the gas and thus lead to a better understanding of the mechanisms
that determine the efficiency of star formation.

High resolution observations also reveal the importance of galactic dynamics. The organization of star
formation by dynamically induced gas flows and dynamical suppression
of star formation becomes evident when observations
reach scales smaller than the characteristic sizes/widths of
dynamical features like bars and spiral arms. As a result, high resolution observations offer insight into how galaxy morphology affects where stars are
born. For example, at high resolution, star-forming regions tend to
appear downstream of the arms seen in gas \citep[e.g.,
  see][]{schinnerer13,egusa17,schinnerer17,kreckel18}. Studying M51,
\citet{meidt13} and \citet{schinnerer13} identified strong variations in
the ratio between the star formation rate and the molecular gas mass \citep[see
  also][]{leroy17a}. Most strikingly, they found regions in the inner
spiral arms that have immense gas reservoirs but little star
formation. \citet{meidt13} attributed these features to dynamical
suppression of star formation \citep[which also manifests in early
  type galaxies at larger scales, e.g.,][]{davis14} rather than the absence
of high density gas \citep[as confirmed by][]{querejeta19}.

To date, most high resolution work has focused on case studies of
individual galaxies. Studies targeting samples of many galaxies have achieved at best a few
hundred pc to $\sim$kpc resolution, and often treated whole galaxies as
a single unit. A clear next step is to systematically examine how
tracers of massive star formation compare to tracers of molecular gas
at high resolution across a large sample.

In this paper, we compare CO emission, tracing molecular gas, to \ha\ emission,
tracing recent star formation, across a large area in eight nearby
galaxies. Our maps all have spatial resolution of $140$~pc, and
often better. This means that an individual resolution element
approaches the size of a large GMC or {\sc Hii} region and that we
expect only one or a few such regions per beam. Our sample shows
diverse morphologies, including spiral arms, bars, nuclear starburst
rings, and flocculent structure. This allows a comparative analysis
and offers the chance to identify general and environmental trends,
which was challenging with previous individual-galaxy case studies.

Specifically, we compare the location of \ha\ emission, a classic
tracer of recent massive star formation \citep[e.g., reviews
  by][]{kennicutt98b,kennicutt12} to the location of CO emission, a
tracer of molecular gas \citep[e.g., reviews by
][]{fukui10,bolatto13,heyer15}. We divide each galaxy into regions where only CO emission is bright, only \ha\ emission is bright, or both tracers are bright. Then, we quantify the fraction of lines of
sight and the fraction of emission from each type of region. We repeat this analysis
for each galaxy for a range of spatial resolutions. We examine the
changing statistics and morphology of the two tracers as a function of
scale. Doing so, we quantify how the coincidence of gas and star
formation observed at low resolution changes as we resolve galaxies at
the scale of an individual star-forming region.

The PHANGS (Physics at High Angular resolution in Nearby GalaxieS\footnote{\url{www.phangs.org}}) 
collaboration is carrying out high physical resolution surveys of molecular gas and star formation tracers in many nearby galaxies.
The observations enabling this analysis are $\sim 1\arcsec$ resolution
CO line emission maps from the PHANGS-ALMA survey \citep[][]{leroy19b},
which is mapping CO emission from the disks of $\sim$80 nearby galaxies with high
fidelity for the first time. This angular resolution matches that of
seeing-limited \ha\ observations and allows $\lesssim 100$~pc
resolution out to distances of up to $\sim 18$~Mpc. For comparison,
previous large CO surveys typically achieved $\sim 10-45\arcsec$ resolution
or targeted only the innermost parts of galaxies with limited
sensitivity.

The paper is organized as follows: The data used are presented in \S
\ref{sec:data}. We describe our methodology in \S \ref{sec:method},
present our results in \S \ref{sec:results}, and discuss the
implications of our findings in \S \ref{sec:discussion}. We summarize
our findings in \S \ref{sec:summary}.


\section{Data}
\label{sec:data}

We require high spatial resolution imaging of the molecular gas and a
tracer of recent star formation on scales of $\lesssim$100\,pc. This
ensures that each individual resolution element approaches the size of
a giant molecular cloud \citep[][]{sanders85} or a massive \hii\ region
\citep[][]{oey03,azimlu11}.

To trace recent massive star formation, we use narrowband \ha+[\nii]
imaging. At the $d <$ 17\,Mpc distance to our targets, seeing-limited
images achieve $\approx1\arcsec\,\lesssim\,100$~pc resolution. We pair
these with $\approx1\arcsec$ resolution low-$J$ CO line imaging to trace
the molecular gas.

We analyze the first seven galaxies targeted by PHANGS-ALMA
\citep[][]{leroy19b}, which also have \ha\ imaging of sufficient
quality for our analysis. Our sample also includes the grand-design
spiral galaxy M\,51 using data from the PAWS survey
\citep[][]{schinnerer13}. We summarize the final sample of eight
galaxies in Table \ref{tab:sample}.

\begin{deluxetable*}{ccccccccll}
\tablecaption{Key Parameters of Sample Galaxies.\label{tab:sample}}
\tablehead{
\colhead{Name} &
\colhead{Distance} &
\colhead{Inclination} &
\colhead{$R_{25}$} &
\colhead{SFR} &
\colhead{$\rm log(M_{\star}$)} &
\colhead{$\rm A_{V,MW}$} &
\colhead{Metallicity} &
\colhead{Hubble Type} &
\colhead{Activity}
\\
\colhead{} &
\colhead{(Mpc)} &
\colhead{(degrees)} &
\colhead{(arcmin)} &
\colhead{$\rm (M_{\odot}~yr^{-1})$} &
\colhead{$\rm (M_{\odot})$} &
\colhead{(mag)} &
\colhead{$\log(\frac{O}{H})_{\rm PP04}$} &
\colhead{} &
\colhead{}
}
\startdata
NGC\,0628    & 9.8  & 9  & 4.9 & 1.83 & 10.2 & 0.192  & 8.51 & Sc-A & H{\sc II} \\
NGC\,3351    & 10.0 & 45 & 3.6 & 1.30 & 10.2 & 0.076 & 8.80 & Sb-B & H{\sc II}; Sbrst\\
NGC\,3627    & 10.6 & 55 & 5.1 & 3.54 & 10.6 & 0.091 & 8.74 & Sb-AB & LINER; Sy\,2\\
NGC\,4254    & 16.8 & 38 & 2.5 & 5.44 & 10.5 & 0.106  & 8.68 & Sc-A & LINER; H{\sc II}\\
NGC\,4321    & 15.2 & 38 & 3.0 & 3.81 & 10.6 & 0.072 & 8.71 & Sbc-AB & LINER; H{\sc II} \\
NGC\,4535    & 15.8 & 41 & 4.1 & 2.22 & 10.4 & 0.053 & 8.78 & SAB(s)c & LINER; H{\sc II} \\
NGC\,5068    & 5.2  & 27 & 3.7 & 0.29 &  9.3 & 0.281  & 8.35 & Scd-AB & -- \\
NGC\,5194    & 8.6  & 21 & 6.9 & 4.91 & 10.7 & 0.096 & 8.70 & SA(s)bc pec & H{\sc II}; Sy2.5
\enddata
\tablecomments{The galaxy parameters are adopted from \citet[][]{leroy19a} where the original references can be found. For 
NGC\,4535, the Hubble type and activity are taken from NED.}
\end{deluxetable*}


\begin{figure*}[tbh]
  \includegraphics[width=\textwidth,angle=0]{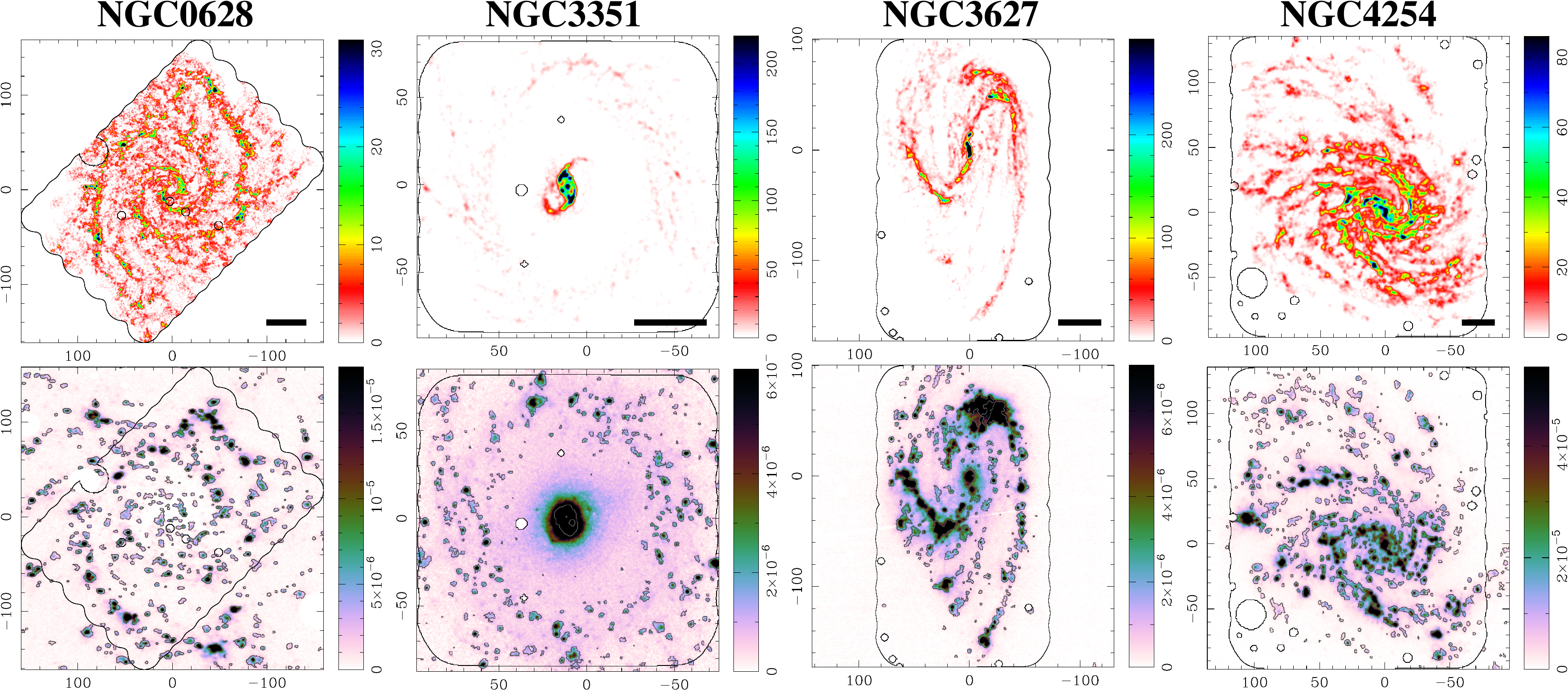}\hfill
  \par \addvspace{0.7cm}
  \includegraphics[width=\textwidth,angle=0]{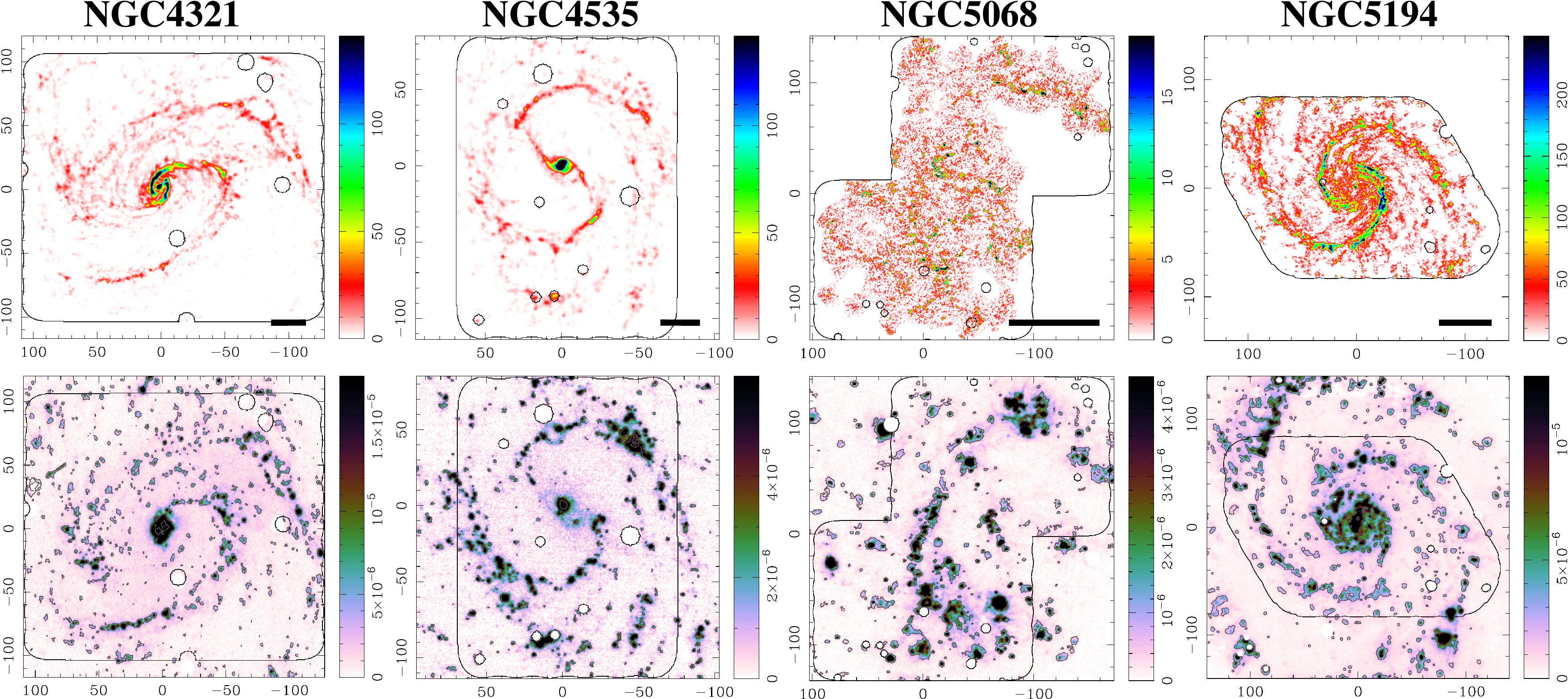}\hfill
\caption{Gallery of CO integrated intensity ({\em first and third
    row}) and H$\alpha$ maps ({\em second and fourth row}) used in
  this paper.  The galaxies shown ({\em from left to right}) are
  NGC\,0628, NGC\,3351, NGC\,3627 and NGC\,4254 ({\em top two rows}),
  and NGC\,4321, NGC\,4535, NGC\,5068 and NGC\,5194 ({\em bottom two
    rows}).  The outer black contour in the CO (\ha) maps indicates
  the edge of our analysis region, i.e., the field that is common to
  both the CO and \ha\ datasets. Empty circles indicate masked
  foreground stars or background galaxies. Note that we use a linear
  color stretch that is different for each galaxy, and for which
  saturation corresponds to the 10$^{\rm th}$ percentile value in each
  map. The CO maps are displayed in units of \kkms. The H$\alpha$ maps, which are corrected for [\nii] contribution
  $f_{\nii}$, \ha\ transmission $T_{\ha}$ and foreground Galactic
  extinction $\rm A_{V,MW}$, are in units of \msolyr. The emission that we identify with
  \hii\ regions is indicated with a gray contour on the \ha\ maps (see
  Section~\ref{subsec:DIG}). The axes are in offset arcseconds from   the reference position of each map. The black bar in the bottom
  right corner of each CO intensity map indicates a spatial scale of 2\,kpc.
\label{fig:gallery_p1}}
\end{figure*} 

\clearpage
\newpage


\begin{rotatetable}
\begin{deluxetable*}{lcccccccccccccc}
\tabletypesize{\tiny}
\tablecaption{Datasets and Parameters used for CO and \ha\ maps.\label{tab:coha_data}}
\tablehead{
\colhead{Name} &
\colhead{FoV$_{\rm CO}$} &
\colhead{$\rm \sigma_{CO}$} &
\colhead{$\rm \theta_{CO}$} &
\colhead{$\rm \theta_{best,CO}$} &
\colhead{$\rm L_{CO}$} &
\colhead{$f_{\rm CO}$} &
\colhead{$\rm \theta_{\ha}$} &
\colhead{$\rm \theta_{best,\ha}$} &
\colhead{$f_{\rm N{\sc II}}$} &
\colhead{$\rm T_{\ha}$} &
\colhead{$\rm L_{\ha, int}$} &
\colhead{SFR} &
\colhead{$f_{\rm DIG}$} &
\colhead{$\log L_{\ha,{\rm min}}$}
\\
\colhead{} &
\colhead{(''$\times$'')} &
\colhead{(\kkms)} &
\colhead{('')} &
\colhead{(pc)} &
\colhead{($\rm 10^8\,K\,km\,s^{-1}pc^2$)} &
\colhead{} &
\colhead{('')} &
\colhead{(pc)} &
\colhead{} &
\colhead{} &
\colhead{($\rm 10^{41}erg\,s^{-1}$)} &
\colhead{($\rm M_{\odot}yr^{-1}$)} &
\colhead{} &
\colhead{${\rm erg\,s^{-1}}$}
\\
\colhead{(1)} &
\colhead{(2)} &
\colhead{(3)} &
\colhead{(4)} &
\colhead{(5)} &
\colhead{(6)} &
\colhead{(7)} &
\colhead{(8)} &
\colhead{(9)} &
\colhead{(10)} &
\colhead{(11)} &
\colhead{(12)} &
\colhead{(13)} &
\colhead{(14)} &
\colhead{(15)} 
}
\startdata
NGC0628  & 319$\times$319\tablenotemark{a} & 0.8 &  1.12  &  60           &   1.85 &  0.72    &    1.73 &  \textbf{90}  &   0.019 &  0.938  & 0.7 &  0.4 & 0.39  &  37.5 \\
NGC3351  & 172$\times$166\tablenotemark{a} & 0.7 &  1.46  &  \textbf{80}  &   0.92 &  0.65    &    1.16 &  60           &   0.209  & 0.972  & 0.6 &  0.3 &  0.56  &  36.5 \\
NGC3627 &  158$\times$273\tablenotemark{a} & 0.9 &  1.57  &  \textbf{90}  &   6.14 &  0.91    &   1.44 &  80          &    0.223 &  0.986  & 1.6 &  0.9 &  0.41  &  36.7  \\
NGC4254 & 199$\times$232\tablenotemark{a}  & 0.4 &  1.71  &  \textbf{140} &  11.88 &  0.86    &   1.52 &  130          &   0.235  & 0.979  & 4.8 &  2.6 &  0.40  &  36.6 \\
NGC4321 & 237$\times$211\tablenotemark{a}  & 0.5 &  1.64  & \textbf{130}  &   5.67 &  0.76    &  1.28 &  100          &   0.112 &  0.838  & 1.6 &  0.9 &  0.58  &  36.5 \\
NGC4535 & 145$\times$199\tablenotemark{a}  & 0.5 &  1.56  & \textbf{120}  &   2.89 &  0.72    &   1.23 &  100          &   0.168  & 0.915  & 0.8 &  0.4 & 0.49  &  36.6 \\
NGC5068 & 269$\times$274\tablenotemark{a}  & 1.3 &  1.00  &  30           &   0.06 &  0.31    &   1.34 &  \textbf{40}  &   0.0770 & 0.993  & 0.3 &  0.2 & 0.35  &  37.0 \\
NGC5194 & 280$\times$180\tablenotemark{b}  & 4.4 &  1.06  &  50           &  10.97 &  0.95    &   1.83 &  \textbf{80}  &   0.228  & 0.983  & 1.4 &  0.7 &  0.47  &  37.6 
\enddata
\tablenotetext{a}{~\bco\ observed by ALMA, \citet[][]{leroy19b}}
\tablenotetext{b}{~\aco\ observed by IRAM, \citet[][]{schinnerer13,pety13}}
\tablenotetext{c}{CTIO\,1.5m, 14.5'$\times$14.5' FoV, CT6586/20 filter, SINGS $\rm 5^{th}$ delivery document\footnote{\url{http://irsa.ipac.caltech.edu/data/SPITZER/SINGS/doc/sings_fifth_delivery_v2.pdf}}}
\tablenotetext{d}{KPNO\,2.1m, 10.2'$\times$10.2' FoV, KP1563 filter, SINGS $\rm 5^{th}$ delivery document}
\tablenotetext{e}{KPNO\,2.1m, 10.2'$\times$10.2' FoV, KP1564 filter, SINGS $\rm 5^{th}$ delivery document}
\tablenotetext{f}{MPG\,2.2m/WFI, 16'$\times$16', 856 filter, \citet[][]{razza19}}
\tablenotetext{g}{CTIO\,1.5m, 10.2'$\times$14.3' FoV, MCELS6568/30 filter, \citet[][]{meurer06}}
\tablecomments{Summary of data products used for the CO and \ha\ emission as well as relevant parameters derived for each data product. The information provided for column:
(1) galaxy name,
(2) size of the area mapped in CO line emission,
(3) characteristic $1\sigma$ sensitivity corresponds to the average RMS across the integrated intensity CO map at the resolution given in column (4),
(4) angular resolution refers to the round Gaussian CLEAN beam used for restoration; except for NGC\,5194 where the geometric mean of the elliptical CLEAN beam is given,
(5) highest native spatial resolution achieved in CO (If the resolution of the CO data sets the limiting spatial scale of our analysis, it is highlighted in bold-face.), 
(6) integrated CO line flux present in the final maps used for our analysis (i.e. after applying a threshold at 12.6\,\mpcsq{} at 140\,pc resolution and artifact mask, within the common FoV),
(7) fraction of integrated CO emission contained in maps used for the analysis (see column (6)) relative to integrated CO emission in the PHANGS-ALMA internal data release v3.3 cube\footnote{
The delivered 0th moment maps have a much better flux recovery ($\gtrsim95$\%). The drop in flux between the delivered maps and our input maps is mostly due to the  CO threshold, which is set by the CO map with the poorest sensitivity at 140\,pc scale (NGC\,5194, $3-\sigma(\Sigma_{mol,140pc}) \sim 12.6$\,\mpcsq)},
(8) angular resolution of the \ha\ maps ($\rm \theta_{\ha}$) as determined on stars in the maps,
(9) highest native spatial resolution achieved in \ha\ (If the resolution of the \ha\ data sets the limiting spatial scale of our analysis, it is highlighted in bold-face.), 
(10) \nii\ fraction $f_{\nii}$ as determined taking the filter response curves into account (see text for details),
(11) transmission correction for \ha\ $T_{\ha}$ as determined taking the filter response curves into account (see text for details),
(12) integrated \ha\ line emission present in the final maps used for our analysis (i.e. within our FoV and with foreground stars masked). 
(13) SFR based on integrated \ha\ line emission (from column (13)) present in map used for analysis,
(14) contribution of DIG emission to total \ha\ emission, and
(15) minimum \hii\ region luminosity, calculated as the \ha\ surface brightness threshold integrated over the beam area, corresponding to our DIG separation strategy.
} 
\end{deluxetable*}
\end{rotatetable}

\clearpage

\subsection{CO Images}
\label{subsec:co_data}

PHANGS-ALMA imaged the CO(2-1) emission in our targets using the
ALMA 12-m, 7-m, and total power antennas. In this paper, we use a
pilot sample observed during Cycle~1 (project 2012.1.00650.S;
NGC\,0628) and Cycle~3 (projects 2013.1.00925.S and 2013.1.00956.S;
remaining galaxies except NGC\,5194). We chose our target area to
match the region of active star formation indicated by an intensity
contour of 1\,MJy\,sr$^{-1}$ in the WISE 22 micron image.

We use integrated intensity maps constructed from the data cubes
delivered in PHANGS-ALMA internal data release 3.3 as our starting
point. The construction of these maps is described in
\citet[][]{leroy19b}, and their properties are summarized in
Tab.\,\ref{tab:coha_data}. Briefly, these were imaged in CASA
5.4.0. They should be sensitive to emission on all spatial scales due
to the inclusion of data from both the 12-m array and the Morita Atacama Compact Array (ACA). Four of the galaxies (NGC\,3627, NGC\,4254,
NGC\,4321, and NGC\,5068) were observed in two separate
large (100+ pointings) mosaics. We match the beams of the two halves via
convolution and combined them via linear mosaicking. This results in
moderately different noise properties for the two halves of the
map. After calibration and imaging, the resulting data cubes were
convolved to have a round beam. The typical RMS noise in brightness
temperature units is $\sigma_{RMS}\,\approx\,0.17$\,K per 2.5\,\kms\ channel,
but varies slightly between cubes.

We use the integrated intensity maps delivered in the internal data
release v3.3. These are constructed by applying a ``broad'' signal
identification mask to the spectral line cube before summing all valid
pixels across the observed velocity range. Specifically, the broad
mask is the logical union of two signal identification masks, the
first generated from the data cube at the best available resolution
and the second from a version of the cube that has been smoothed to a
spatial resolution of 500\,pc. In both cases, we identify emission
above $3.5\sigma$ in 3 consecutive channels of the cube, and expand it
to include contiguous regions in ($x,y,v)$ space that show emission
above $2\sigma$ over 2 consecutive channels. The resulting ``broad''
integrated intensity maps have high completeness, meaning that they
include most CO emission in the cube, even emission present at only
modest signal-to-noise. More quantitatively, the flux recovery in
the v3.3 broad integrated intensity maps (which are our input maps for the analysis) varies between 93\% (NGC\,4321) and 99\% (NGC\,3627) of the
total CO flux that is present in the input best resolution data cube
\citep[see also discussion in][]{sun18}. Sensitivity estimates for
these integrated intensity maps are tabulated in
Tab.\,\ref{tab:coha_data}. These are spatial averages of the
sensitivity across each map, calculated as $\sigma_{\rm RMS} \times
\sqrt{N} \times \Delta v$, where $N$ is the number of channels in the
signal identification mask for that sightline, and $\Delta v$ is the
channel width. The equivalent inclination-corrected 1$\sigma$
integrated molecular mass surface density sensitivity limits vary between 2.5 and
7.4\,$\mpcsq$ across the PHANGS galaxies, assuming a standard Galactic
conversion factor of $\alphaco=4.35$\,\acou\ \citep{bolatto13} and a
\bco\ to \aco\ brightness temperature ratio of $R_{21}=0.7$.

For NGC\,5194 (M51), we use the $1\arcsec$ resolution integrated
CO(1-0) image from PAWS \citep[PdBI Arcsecond Whirlpool
  Survey,][]{schinnerer13}. PAWS combined observations from the IRAM
PdB interferometer and 30m single dish telescope, ensuring that the
map includes emission on all spatial scales
\citep[see][]{pety13}. Details of the observations and data reduction
are presented in \citet[][]{pety13}. The PAWS map has an average
$1\sigma$ sensitivity of $\sim4.4$\,\kkms, corresponding to an integrated mass
surface density sensitivity of $\sim19$\,$\mpcsq$.

\subsection{\ha\ Images}
\label{subsec:Ha}

We use narrowband \ha\ imaging to trace recent massive star
formation, drawing from a mixture of new and literature data. For the literature data, we adopt the
image that best matched the following criteria: (i) field-of-view
(FoV) sufficient to cover the CO data, (ii) angular resolution similar
to the CO data, and (iii) a corresponding R band image observed with
the same telescope under similar conditions. For several galaxies where no literature data matched our criteria, we obtained new Wide-Field Imager (WFI) narrowband and R band imaging using the
MPG/ESO\,2.2m telescope. Details on the datasets used can be found in Tab.\,\ref{tab:coha_data}.

Specifically, we used data from the SINGS ancillary survey \citep[][]{kennicutt03}
for NGC~0628, NGC~3351, NGC~4254, NGC~4321, and NGC~5194. Details of
the observations, including the associated R band data used for
continuum subtraction, can be found in the SINGS fifth delivery
document\footnote{\url{http://irsa.ipac.caltech.edu/data/SPITZER/SINGS/doc/sings_fifth_delivery_v2.pdf}}. 
The continuum R band data reach a depth of 25
mag/arcsec$^2$ with a signal-to-noise of $\sim$\,10. The
\ha\ narrowband filter data were observed for 1800\,s.

NGC\,5068 was observed in the \ha\ narrowband and R band as part of
the Survey for Ionization in Neutral Gas Galaxies \citep[SINGG;
][]{meurer06} at the CTIO. 
NGC\,5068 was also observed for 1800\,s in the
narrowband, and 360\,s in the R band.

For NGC\,3627 and NGC\,4535, the literature data were not of
sufficient quality to match our criteria. Therefore, we use data from
a new survey of PHANGS targets using the WFI
instrument on the MPG/ESO\,2.2-m telescope at La Silla
Observatory. This survey obtained both narrowband \ha\ 
and Rc band 
using WFI. 
The galaxies were placed in a
single CCD, avoiding chip gaps, allowing an uninterrupted FoV of
16'$\times$16'. These data were astrometrically and photometrically
calibrated using GAIA DR2 catalogs cross-matched to all stars in the
full FoV of the WFI images. Full details of this survey, including
observations, reduction, calibration and sky subtraction can be found
in \cite{razza19}.

\subsubsection{Sky subtraction}
\label{subsubsec:skysubtraction}

We subtract a sky background from the SINGS \ha\ data. To do this, we
first mask all bad pixels and bright sources. Then, we mask all
emission that exceeds the median of the remaining image by $>3$ times
the rms noise. We convolve this mask by $\sim\,3\times{\rm FWHM}$ to
further mask out any diffuse light from bright sources or the
galaxy. We then fit and subtract a plane from the masked image. In all
cases, our targets fill a significant fraction of the image. In a few
cases this caused the plane fit to fail. In these cases, the sky
background was set to be the median of the masked image. The typical
sky brightness was 20-21 mag/arcsec$^2$ in the R band images.

For NGC\,5068 and our WFI data, the images are already sky subtracted
by \citet{meurer06} and \cite{razza19}, respectively. We repeated the
above analysis and confirm that the residuals appear consistent with
noise.

\subsubsection{Seeing and astrometry}
\label{subsubsec:seeing}

After sky subtraction, we fit all point sources in both the \ha\ and R
band images to determine the average seeing\footnote{This was done
  using the IDL implementation of the DAOPHOT routines \texttt{APER}
  and \texttt{GETPSF}:
  \url{https://idlastro.gsfc.nasa.gov/ftp/pro/idlphot/aper.pro}.}. When
the seeing differed significantly between the R band and \ha\ images
($>\,0.5$ pixel), we convolved the higher resolution data to match the
lower resolution. Typical seeing for the data range from 1\farcs2 to
2\farcs1 ~(see Tab.~\ref{tab:coha_data} for
individual galaxies).

For the SINGS and SINGG data, the astrometry was already confirmed by
matching stellar sources to the US Naval Observatory A2.0 database
\citep[][]{monet98}. \citet[][]{meurer06} matched $\sim\,100$ and
found an accuracy of 0\farcs4, similar to the 0\farcs5 found for the
SINGS data.

We require accurate absolute astrometry to assess spatial offsets
between \ha\ and CO emission. Therefore, we used Gaia DR2
\citep{gaia16,gaia18} to further refine the absolute astrometry. We
fit $\sim\,50$ stars per R band image with an rms of
$\sim$\,50--60\,mas on the required offsets which we applied. This yielded a final astrometric accuracy of
0\farcs1--0\farcs2. 
The astrometry of the WFI images was calculated using the Gaia DR2 catalog directly and so already reach this same level of accuracy.

\subsubsection{Continuum subtraction}
\label{subsubsec:continuumsubtraction}

Once the \ha\ and R band images are aligned and at the same
resolution, we determine the scale of the R band continuum in the
\ha\ narrowband image. If the images are correctly flux calibrated,
this scale should be one. However to confirm this we determine the
median flux ratio of the matched non-saturated stars between the
narrowband and R band images.

Using this flux ratio as a basis, we obtain a first estimate of the
\ha\ flux by subtracting the R band image from the
narrowband. However, the \ha+N{\sc ii} line also contributes to the R band
data. Therefore, using the estimated \ha+N{\sc ii} image we determine the
\ha+N{\sc ii} contribution to the R band image. We subtract this estimated line
contamination from the R band image and iterate this process until
successive continuum estimates differ by less than $1\%$. Then we
subtract this continuum estimate to achieve a flux-calibrated line
image.

Note that this approach does not account for a varying continuum slope
in the R band. We might expect some variation in the narrowband
continuum to R band continuum ratio due to, e.g., a variation in the
stellar population across the galaxy disk (Razza et al. in prep.). 

\subsubsection{Correction for transmission, [N{\sc ii}] contamination, and Galactic extinction}
\label{subsubsec:transmission_hii_extinction}

We correct for transmission loss of the \ha\ emission line, which can
be red-shifted out of the chosen narrowband filter, and for the
contribution of the [\nii]$\lambda6548,6583$ emission lines, which may
fall within the narrowband filters.

To do this, we use an integrated, high signal-to-noise H{\sc ii}
region spectrum from the MUSE observations of NGC\,0628 presented in
\citet[][]{kreckel16}. We treat this as a characteristic spectrum for
all of our targets. Our targets are all relatively massive, relatively
face-on spiral galaxies, so we expect the NGC\,0628 spectrum to be
reasonably representative of the typical H{\sc ii} region in our
sample. The NGC\,0628 template spectrum has an intrinsic [N{\sc
    ii}]/\ha\ ratio of 0.3. This is close to, but slightly lower than
some of the global estimates for the [N{\sc ii}]/\ha\ in our targets
from, e.g., \citet[][]{kennicutt11}. We note possible variation in
[N{\sc ii}]/\ha\ as a source of uncertainty.

We shift this template spectrum to the redshift of each galaxy. Then,
using the transmission curves for the filters listed in the notes of
Tab.\,\ref{tab:coha_data}, we calculate the contribution of the
\ha\ and [N{\sc ii}] lines to the narrowband image (see
Tab.\,\ref{tab:coha_data}). Based on this calculation, we subtract the
likely contamination by [N{\sc ii}] to create an \ha -only surface
brightness image ($F_{\ha}=F_{\rm line}(1-f_{[\nii]})/T_{\ha}$).

As a final step, we correct all images for foreground Galactic
extinction using the map of \citet[][]{schlafly11}.

\subsubsection{Filtering out emission from diffuse ionized gas (DIG)}
\label{subsec:DIG}

Our analysis uses \ha\ emission to pinpoint the spatial location of
recent high-mass star formation, i.e., we are specifically interested in \ha\ emission arising from H{\sc ii} regions that surround massive stars. 
In this context, \ha\ emission arising from diffuse ionized gas (DIG) must be excluded from the analysis. This \ha\ emission does not originate from gas that is ionized locally by young massive stars. Unfortunately, DIG can contribute 50\% or more to the observed \ha\ flux of a galaxy \citep[e.g., review by][]{haffner09}. The origin of the ionizing photons producing the DIG is still debated, with some or indeed all of these photons originating from massive young stars \citep[e.g., due to Lyman-continuum leakage from \hii\ regions; ][]{weilbacher18}. For our purposes, the key issue is that the DIG emission is not necessarily co-spatial with the young stellar population powering the H{\sc ii} regions \citep[see, e.g., ][]{kreckel16}. Instead DIG generally exhibits a more smooth appearance. 

There are two main methods to separate the \ha\ emission arising from H{\sc ii} regions and the gas associated with the DIG. The first method, based on morphology, decomposes the \ha\ maps into bright knots and a diffuse background \citep[e.g.,][]{thilker02,oey07} using for example the software {\sc Hiiphot} \citep{thilker00}. Alternatively, optical spectroscopy can be used to identify emission arising from DIG, which is warmer and less dense than the H{\sc ii} region gas \citep[e.g. ][]{blanc09,haffner09,kaplan16,tomicic17}.

After correcting for Galactic extinction, we attempt to remove the DIG contribution from our \ha\ maps. We use the following unsharp masking scheme which is qualitatively similar to the approaches mentioned above. We implemented a two step approach which is first identifying diffuse emission on scales larger than H{\sc ii} regions, and in a second iteration attempting to take into account variations in the strength of the DIG contribution on the scales of galactic structures such as spiral arms. The filtering scales and gain represent tuning parameters informed by physical expectations:

\begin{enumerate}
\item {\bf Unsharp mask with a 300~pc kernel:} We smooth our original image with a Gaussian kernel with FWHM 300~pc. We choose 300~pc because we do not expect {\sc Hii} regions to reach sizes much larger than that scale \citep[e.g.,][]{oey03,azimlu11,whitmore11}. Then, we subtract this smooth version of the image from the original image. We identify likely {\sc Hii} regions as the parts of the map still detected at high signal-to-noise in this filtered map.

\item {\bf Subtract a scaled version of the initial {\sc Hii} regions from the DIG map:} We subtract a scaled version of the {\sc Hii} regions identified in the previous step from the original map. The scaling factor, which is $0.33$, reflects that this is an initial estimate and that we do not want to oversubtract at this stage.

\item {\bf Unsharp mask with an 750~pc kernel:} We smooth our {\sc Hii} region subtracted image with a Gaussian kernel that has FWHM
  750~pc. This scale is large enough to typically encompass galactic structures, such as spiral arms, that might show higher levels of DIG contribution and clustering of {\sc Hii} regions
\citep[e.g.][]{kreckel16}. Then, we subtract this smooth version of the image from the \textit{original} image. We identify our final set of {\sc Hii} regions as the parts of the map still detected at high signal-to-noise in this filtered map.

\end{enumerate}

We proceed by blanking all regions outside the identified {\sc Hii} regions and treating the masked map as our best estimate for the location of massive star formation. 
The final \ha\ maps used for our analysis are constructed by applying this mask to the calibrated and corrected \ha\ map.
The flux in this masked map represents our best estimate of the relative intensity of star formation, because we expect most of this \ha\ emission to arise from gas photo-ionized by massive stars (rather than, e.g., shocks). On average, this process removes $\sim 50\%$ of the \ha\ emission from the initial maps.

The tuning parameters that we adopted were optimized to yield a procedure that identifies bright knots of \ha\ emission, which are likely {\sc Hii} regions, when applied to our whole sample. Three of the authors independently inspected the resulting masks to
confirm that they indeed isolate the bright knots in each of our
\ha\ maps, without subdividing the emission from large {\sc Hii}
regions. In a few targets, we further confirmed that our approach agrees sufficiently well with masks based on {\sc Hiiphot} and with spectroscopic separation of DIG and {\sc Hii} regions \citep[][]{kreckel16,blanc09}. We experimented with fixed surface brightness cuts to separate the \ha\ emission into the DIG and the \hii\ regions, but found the resulting masks to be much less satisfactory. 
This is likely linked to the fact that the \ha\ luminosity of H{\sc ii} regions depends on the exact physical conditions present in the ionized gas such as ionizing flux, gas geometry etc. which can vary significantly between and within galaxies at our resolution. Lastly, we note that the vertical extent of the DIG can be significant reaching typical scale heights of 1\,kpc \citep[e.g. ][]{rossa03,levy19}.

We note one important additional caveat about our strategy for DIG
removal. When identifying bright regions during the unsharp masking
steps, we use a signal-to-noise threshold. Because the noise and
native resolution of the input \ha\ data vary, the effective
\ha\ surface brightness threshold applied to our fiducial maps ranges
from $I_{\rm H\alpha} [{\rm erg\,s^{-1}\,cm^{-2}\,sr^{-1}}] = 2.21 \times10^{-6}$
(NGC\,4254) to $9.46\times10^{-6}$ (NGC\,5068), corresponding to SFR surface density
limits between 0.0014 and 0.0036\,\msol\,$\rm yr^{-1}\,kpc^{-2}$. For a point source at the native resolution of our data, these sensitivity limits
correspond to \hii\ region luminosities between $\log_{10} L_{\rm
  H\alpha} [{\rm erg\,s^{-1}}] = 36.5$ (NGC\,4321) and 37.6
(NGC\,5194). These limits are comparable to the peak of the
\hii\ region luminosity function measured by narrowband \ha\ imaging
in the literature \citep[e.g.][]{bradley06,oey07}, although the
current generation of optical IFUs are able to detect \hii\ regions
with lower luminosities \citep[e.g.][]{kreckel16,rousseau18}.

\subsubsection{No correction for internal extinction}
\label{subsec:halpha_internal_AV}

Our final \ha\ line flux images trace emission from \hii\ regions
without any correction for attenuation. We expect our images to do a
good job tracing the location of massive star formation, but the maps
may miss the most heavily extinguished regions. Because our analysis focuses
mostly on the location (rather than the amount) of recent massive star
formation, we expect this to be a secondary concern. Based on
literature work in M\,51 \citep[][]{scoville01,schinnerer17}, we do
not expect that such highly extinguished regions occupy a significant
fraction of sightlines in the disks of our galaxies. For our analysis
based on flux, the underestimation of flux is likely stronger for the
regions co-located with molecular gas, where dust is likely well-mixed
with the gas, than for regions where no gas is present.

\section{Methodology}
\label{sec:method}

We aim to quantify the coincidence of molecular gas, traced by CO, and
high-mass star formation, traced by \ha, across a range of spatial
scales. To do this, we adopt the following approach:

\begin{enumerate}

\item \textbf{Threshold the \ha\ images:} The {\bf removal of the significant} DIG contribution to the
  \ha\ images acts as a natural threshold (see \S\,\ref{subsec:DIG}).
  We calculate the total flux in the thresholded image at this stage.
  
\item \textbf{Threshold the CO images:} We clip the CO images at our
  best matching resolution of 140\,pc using a physical threshold of
  12.6\,\mpcsq\ (accounting for galaxy inclination) or roughly an $\rm A_V\approx1$. This corresponds
  to the equivalent $3\,\sigma$ mass surface density sensitivity of
  our CO map with the lowest sensitivity at this spatial scale
  (NGC\,5194).

\item \textbf{Convolve to coarser physical resolutions:} We convolve
  each thresholded image to a succession of coarser resolutions,
  ranging from our best matching resolution of 140\,pc to $1.5$~kpc
  (to match to previous studies working at coarser resolution).

\item \textbf{Clip the low intensity emission in the convolved
  images:} For each convolved image, we blank the faintest lines of
  sight that collectively contribute $2\%$ of the total flux in the
  image. This suppresses convolution artifacts.

\item \textbf{Measure the coincidence of the two tracers at each
  resolution:} We divide each galaxy at each resolution into regions
  with visible CO only, visible \ha\ only, or both CO and
  \ha\ emission present. We measure the fraction of sightlines and the
  fraction of flux in each region type at each resolution.

\end{enumerate}

This approach represents a specific application of a more general
methodology, which would be to measure the joint distribution
functions of CO and \ha . In the rest of this section, we motivate the
specific choices that we have made. When describing the results in the next sections, we
 use the words 'sightlines' and 'pixels' interchangeably.

\subsection{Thresholding and total flux estimation}
\label{subsec:fluxthresh}

We threshold both the CO and \ha\ images before comparing them. This
fixes the total flux used in the analysis across scale. It also
ensures that we have high confidence that all emission entering the
analysis corresponds to real high-mass star formation or massive
concentrations of molecular gas.

The total CO and \ha\ fluxes after thresholding within the common FoV
of our 140\,pc resolution images are tabulated in
Tab.\,\ref{tab:coha_data}. For \ha, our threshold is set by our DIG
removal strategy (see \S\,\ref{subsec:DIG}).
For CO, we impose a mass surface density
threshold that allows us to work exclusively with securely detected
pixels across all our CO maps. 
For the CO emission, we do not impose any filtering analogous to what we used to subtract the DIG (from the \ha\ maps). Our reasoning is that while ``diffuse'' CO emission has been identified in galaxies \citep[e.g.,][]{pety13,caldu-primo15,roman-duval16}, the physical nature of this gas remains unclear. Extended CO emission still does indicate the presence of molecular gas, which may very well be in marginally bound clouds and available for star formation. This is not symmetric to the DIG, which represents emission not associated with local massive star formation. This is an important topic for future exploration, but in the interests of maintaining a simple, physically-motivated, and reproducible approach, we do not filter the CO emission. 

For reference, our adopted CO threshold is roughly equivalent to a cut of $\rm A_V\approx1\,mag$ \citep[e.g.,][]{bohlin78}. This is close to the limit of $\rm A_V=2\,mag$ used by \citet{heiderman10} to define the boundaries of local molecular clouds in highly resolved maps. That is, our CO threshold already identifies only pixels where the emission averaged over a 140~pc beam matches that seen within highly resolved local GMCs.

Typically, this threshold retains about 75\% of the total flux in the input CO data cube, but this drops to 30\% in the case of the low-mass galaxy NGC\,5068.

\subsection{Spatial scales, convolution, and clipping}
\label{subsec:conv_and_masking}

We compare the distributions of CO and \ha\ at a series of common
spatial resolutions up to $1.5$~kpc. The best physical resolution
shared by our entire sample is 140\,pc, which is set by the resolution
of our CO map of NGC4254. We adopt 140\,pc as the fiducial measurement
scale for our analysis.

To convolve the data, we smooth the thresholded images using Gaussian
convolution kernels. For each convolved map, we sort the map pixels by
intensity (from brightest to faintest), and reject the faintest pixels
that together contribute 2\% of the total flux. We verified that this
additional clipping level is sufficient to remove faint halos that are
convolution artifacts.
This step sets a large amount of area to have zero intensity at each
scale without significantly affecting the flux in the analysis. We
tested that other reasonable rejection criteria (between 1 and 5\%) do
not affect our conclusions.

Finally, we note that while we apply the same methodology to the CO and \ha\ maps, the output depends on the
intrinsic distribution of the
emission in the input map. Faint emission located next to a
bright peak in the input map will tend to be recovered in the 
lower resolution map, for example, while isolated faint emission is
more likely to fall beneath the 2\% clipping threshold in the lower resolution maps. In this sense, our smooth-and-clip strategy is likely to impact our CO and \ha\ maps differently. Our method for DIG
removal often results in high-brightness regions being surrounded by
zero-valued pixels. On the other hand, the mass surface density threshold
applied to construct our CO maps tends to yield contiguous regions of
relatively faint emission. We note this as another potential area for
improvement.

\subsection{Measurements}
\label{subsec:output}

Using the pairs of matching resolution CO and \ha\ maps, we classify each pixel into one of four categories: 

\begin{itemize}
\item {\bf \em CO only} -- only CO emission is present, \ha\ emission is not detected at this position and scale.
\item {\bf \em \ha\ only} -- only \ha\ emission is present, CO emission is not detected at this position and scale.
\item {\bf \em Overlap} -- both CO and \ha\ emission are present.
\item {\bf \em Empty} -- neither CO nor \ha\ emission are present.
\end{itemize}

\noindent This classification is scale dependent, e.g., a pixel may be
classed as {\em empty} at high resolution, {\em CO only} at
intermediate resolution, and {\em overlap} at low resolution. At high
resolution, a non-negligible fraction (on average around 60-70\%) 
of pixels falls in the last
category (see Fig.\,\ref{fig:allgals_olap_gallery_140pc}), indicating
that large areas of galaxy disks are not directly filled with
star-forming ISM. At coarse resolution, the number of empty pixels
decreases significantly, giving the impression that the star-forming
ISM is ubiquitous. The differences among distributions of empty pixels among
galaxy disks is thus a potentially interesting diagnostic of ISM
evolution and galaxy morphology. In this paper, however, we are most
interested in the impact of host galaxy properties and observing scale
on the relative configuration of the molecular and ionized gas phases,
so we do not analyse the fraction of empty pixels in detail.

At each spatial scale after removing the empty pixels, we calculate the fraction of pixels in each of
the first three categories, i.e. with a detection of CO, \ha\ or
both. At fixed scale, these fractions may be plotted as a pie chart
(Section~\ref{subsec:2d-maps_140pc}). The evolution of these fractions
as a function of scale can be illustrated using a bar graph
(Section~\ref{subsec:2Dmaps_scale}).

To assess the uncertainty on our measurements, we repeated our
analysis while varying the thresholds for CO and \ha\ flux. We tested
varying the CO threshold between 10 and 15\mpcsq\ at 140\,pc scale,
and thresholding at the native resolution of the CO maps using
 signal-to-noise criteria (i.e. 3 and 5$\sigma$ cuts in
integrated intensity). We also varied the unsharp masking parameters
used for DIG removal, varying the size of the smoothing kernels by
25\%, the gain in the first subtraction step by 25\%, and the
signal-to-noise threshold of the mask by $\pm1\sigma$. We adopt the
minimum and maximum values of the pixel fractions obtained from this
suite of tests as an indication in the uncertainty of our
measurements.

\section{Results}
\label{sec:results}

Following the methodology in \S \ref{sec:method}, we measure the
fraction of detected sightlines with \ha\ emission, CO emission, or
both at each spatial scale. For plots in this section, sightlines with
{\em \ha\ only} emission are shaded with red tones, sightlines with
{\em CO only} emission are shaded with dark blue tones, and sightlines
with overlapping CO and \ha\ emission ({\em overlap}) are shaded with
lavender tones.

\subsection{CO and \ha\ at 140 pc resolution}
\label{subsec:2d-maps_140pc}

\begin{figure*}[tbh]
  \centering
\includegraphics[width=\textwidth,angle=0]{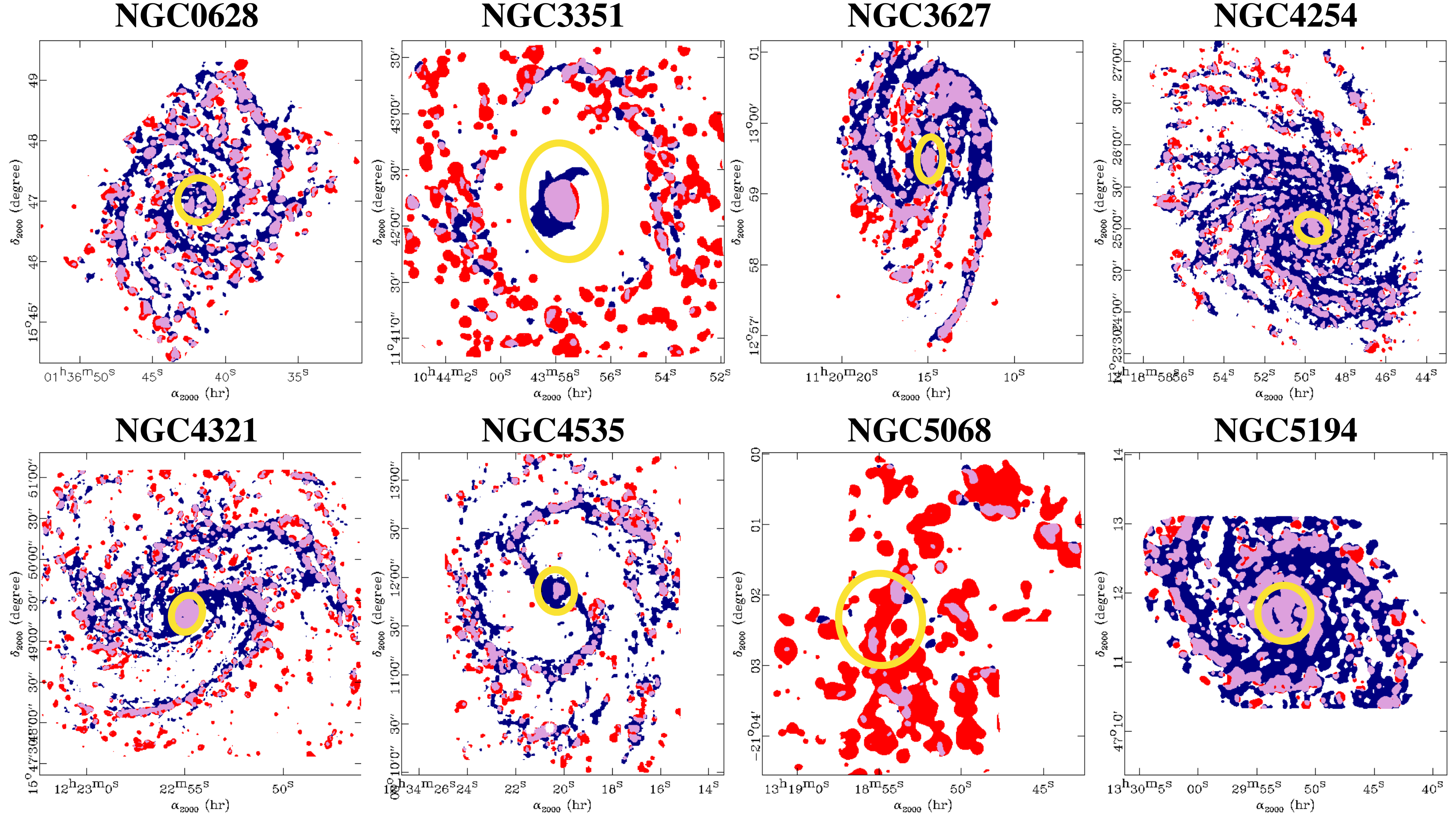}\hfill
\caption{Galaxy maps showing regions with {\em overlapping} CO and
  \ha\ emission (lavender), {\em CO only} emission (dark blue) and
  {\em \ha\ only} emission (red) at a spatial resolution of 140\,pc for our
  sample. The region that we define as the galaxy center is indicated
  in each panel as a yellow ellipse.}
\label{fig:allgals_olap_gallery_140pc}
\end{figure*}

\begin{figure*}[tbh]
  \centering
\includegraphics[width=\textwidth,angle=0]{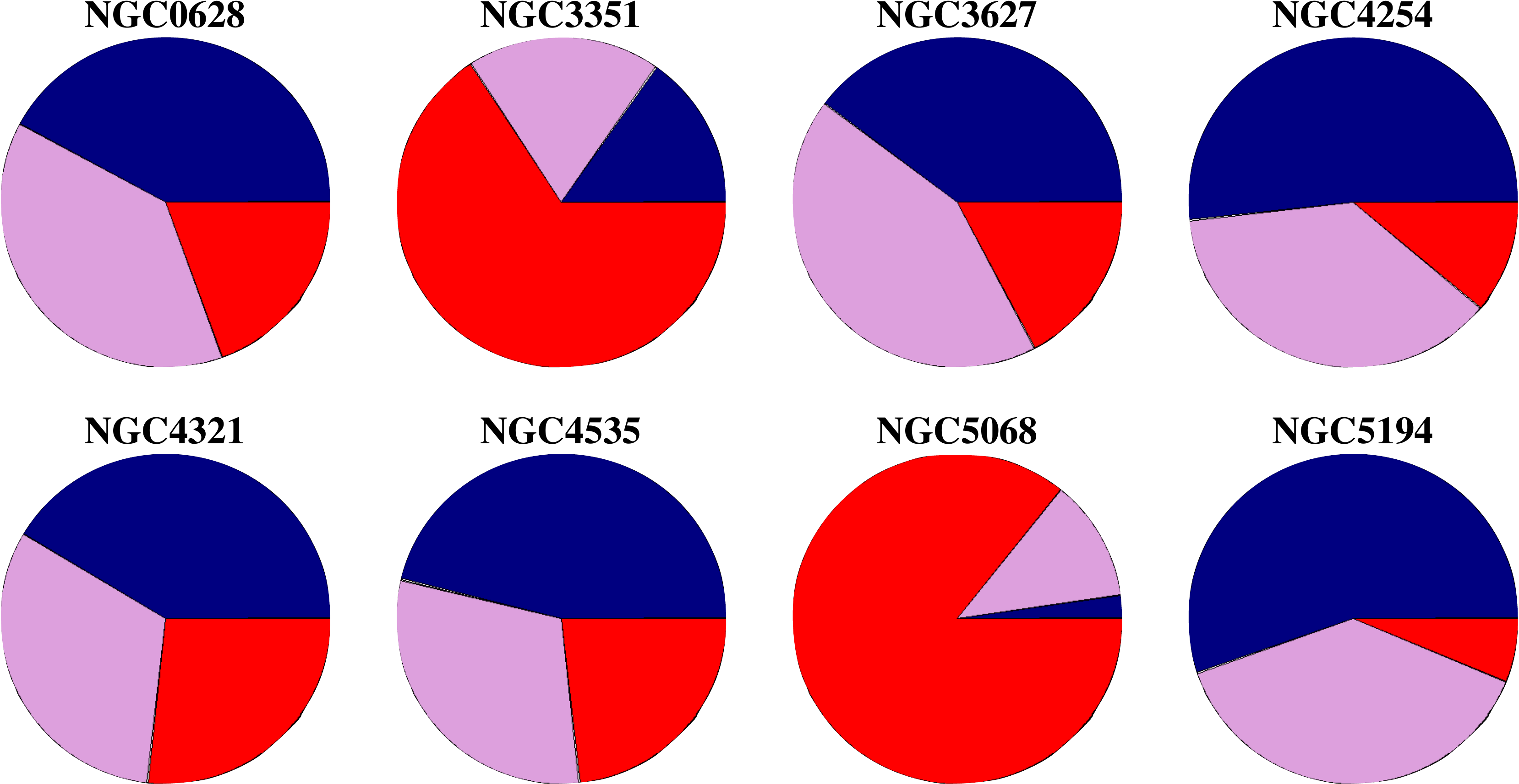}\hfill
\caption{Pie charts indicating the fraction of sightlines with CO and \ha\ emission 
for the 140\,pc resolution maps of our eight galaxies (from {\em left} to {\em right}: 
NGC\,0628, NGC\,3351, NGC\,3627 and NGC\,4254, {\em (top row)}; 
; and NGC\,4321, NGC\,4535, NGC\,5068, NGC\,5194 {\em (bottom row)}).
The color coding is dark blue -- {\em CO only}, red -- {\em \ha\ only}, lavender -- {\em overlap}.
\label{fig:piechart_gallery_p1}}
\end{figure*}

\begin{deluxetable}{c|ccc}
\centering
\tablecaption{Line of Sight Fractions at 140\,pc Scale.\label{tab:piechart}}
\tabletypesize{\scriptsize}
\tablehead{
\colhead{Name} &
\colhead{CO only} &
\colhead{\ha\ only} &
\colhead{overlap} 
\\
\colhead{} &
\colhead{(\%)} &
\colhead{(\%)} &
\colhead{(\%)}
}
\startdata
NGC0628 &    42$^{54}_{27}$ &   19$^{32}_{12}$  &  38$^{42}_{34}$ \\
NGC3351 &    15$^{35}_{8}$ &   66$^{81}_{63}$  &  19$^{24}_{13}$ \\
NGC3627 &    40$^{47}_{33}$ &   17$^{25}_{10}$  &  43$^{49}_{37}$ \\
NGC4254 &  52$^{67}_{39}$ &   11$^{17}_{6}$  &  37$^{43}_{28}$ \\
NGC4321 &   41$^{53}_{29}$ &   27$^{40}_{18}$  &  32$^{38}_{23}$ \\
NGC4535 &   46$^{61}_{37}$ &   23$^{38}_{13}$  &  31$^{37}_{26}$ \\
NGC5068 &    2$^{7}_{1}$ &   86$^{90}_{73}$  &  12$^{20}_{10}$ \\
NGC5194 &   55$^{59}_{52}$ &   6$^{8}_{4}$  &  38$^{39}_{35}$  \\
\hline
Median  &  42  &  23  &  37  \\
Mean  &  37  &  32  &  31  \\
\enddata

\tablecomments{Fraction of sightlines with CO emission only,
  \ha\ emission only, and both CO and \ha\ emission. These fractions
  represent the numbers corresponding to the pie charts shown in
  Fig.\,\ref{fig:piechart_gallery_p1}. The sub- and super-scripts
  correspond to the minimum and maximum values obtained when different
  thresholds for positive detection of CO emission and DIG suppression
  are used (see text). The mean (median) values reported in the final
  row are the means (medians) of our fiducial measurements reported
  for a given sightline category. 
  Note that the medians are not re-normalized, thus the sum will not add up to 100\%.
  }
\end{deluxetable}

Fig.\,\ref{fig:allgals_olap_gallery_140pc} shows results for all
galaxies at 140\,pc resolution in map form. Table \ref{tab:piechart}
summarizes the statistics for these maps, and
Fig.\,\ref{fig:piechart_gallery_p1} visualizes these statistics as pie
charts. Following \S \ref{sec:method}, the error bars in Table
\ref{tab:piechart} show the effect of varying our CO and
\ha\ detection thresholds across their plausible ranges.

At this resolution, emission from one or both of these tracers is
typically present across 50\% or more of our survey area. Of the
detected sightlines, {\em CO only} sightlines (dark blue) are the most common 
detections in more than half of our targets (namely
NGC\,0628, NGC\,4254, NGC\,4321, NGC\,4535, NGC\,5194). In two targets
{\em \ha\ only} sightlines dominate (NGC\,3351 and NGC\,5068). NGC\,3627 -- the most inclined galaxy in our sample with $i\sim 55^{\circ}$ -- is our only target where sightlines with
{\em overlapping} CO and \ha\ emission are the most common.

Across the sample, the median fraction of detected sightlines that
contain {\em CO only} emission is $\sim$40\%. This is comparable to the
fraction of {\em overlap} sightlines ($\sim40$\%), while the fraction of
{\em \ha\ only} sightlines is typically only $\sim20$\%. The clear separation of the
CO and \ha\ emission at our high working resolution (i.e., the lower
overlap fraction compared to sightlines with one tracer only), and the
detection of widespread CO emission above 12.6\,$\msol\,pc^{-2}$ without
associated \ha\ emission (typically {\em CO only} sightlines are more
than 2$\times$ more abundant than {\em \ha\ only} ones) are two main
results of this paper.

Table \ref{tab:byflux_bynum} (see also Fig.\,\ref{fig:appendix_summary}) shows the
results when we consider how the CO and \ha\ {\it fluxes} (rather than the number of
sightlines) are distributed between different categories, i.e. instead of the number of pixels in each category, we consider their respective contribution to the total CO or \ha\ flux. 
On average, we find that $\sim 40\%$ of the CO flux at
140\,pc resolution remains unassociated with bright
\ha\ emission. This is lower than the $\sim 60\%$ of {\it CO only}
sightlines, but implies the same qualitative conclusion that there is
a significant mass fraction of CO-traced molecular gas without
\ha\ emission.

We find a greater discrepancy between the distributions of flux and
sightlines for \ha . At our fiducial 140\,pc scale, the median {\em overlap} fraction for \ha\ emitting sightlines is about two-thirds across our sample. These same regions contribute a median of $\sim80$\% of the total \ha\ flux. These results indicate that these {\em overlap} sightlines, which show both CO and \ha ,
tend to have higher \ha\ intensity than sightlines probing {\em
  \ha\ only} emission. This agrees with a visual inspection of the maps.
Since these {\em overlap} regions are by definition co-spatial with CO-emitting molecular gas, they likely suffer from dust attenuation that we do not account for in the processing of our \ha\ maps (c.f. \S\,\ref{subsubsec:transmission_hii_extinction} and \ref{subsec:halpha_internal_AV}). The true flux contribution of these regions is thus probably higher than the $\sim80\%$ that we measure.

Galaxy centers can contribute significantly to the total flux while
covering only a few sightlines. Fig.\,\ref{fig:allgals_olap_gallery_140pc} and Fig.\,\ref{fig:gallery_p1} show that the galaxy centers in our sample tend to have 
{\em overlapping} high intensity CO and \ha\ emission. In \S\,\ref{appendix:piechart_center}, we
test the degree to which the centers drive the discrepancies between our flux-
and sightline-based measurements. To do this, we repeated the
calculation considering only the main galaxy disks, i.e., excluding the
region within a radius of 1\,kpc of the reference position listed in
Table~\ref{tab:sample} for all galaxies.
For NGC\,3351, we follow \cite{sun18} who adopted a larger radius of 1.5\,kpc
so that the visually distinct inner disk is entirely inside the central
region
of NGC\,3351.
The regions that we define as
central are indicated by yellow ellipses in
Fig.~\ref{fig:allgals_olap_gallery_140pc}. Excluding the
central regions from our analysis improves the agreement between the flux-
and sightline-based measurements for NGC\,3351, but does not significantly alter our
results for other galaxies.

\begin{deluxetable*}{c|cc|cc|cc|cc}
\centering
\tablecaption{Line of Sight and Flux Fractions per Tracer at 140\,pc Scale.\label{tab:byflux_bynum}}
\tablehead{
\colhead{} &
\multicolumn{4}{|c|}{By Number}  & 
\multicolumn{4}{c}{By Flux} 
\\
\colhead{} &
\multicolumn{2}{|c|}{CO Sightlines}  & 
\multicolumn{2}{c|}{\ha\ Sightlines} &
\multicolumn{2}{c|}{CO Flux}  & 
\multicolumn{2}{c}{\ha\ Flux} 
\\
\multicolumn{1}{c|}{Name} &
\colhead{CO only} &
\multicolumn{1}{c|}{CO overlap} &
\colhead{\ha\ only} &
\multicolumn{1}{c|}{\ha\ overlap} &
\colhead{CO only} &
\multicolumn{1}{c|}{CO overlap} &
\colhead{\ha\ only} &
\colhead{\ha\ overlap} 
\\
\multicolumn{1}{c|}{} &
\colhead{(\%)} &
\multicolumn{1}{c|}{(\%)} &
\colhead{(\%)} &
\multicolumn{1}{c|}{(\%)} &
\colhead{(\%)} &
\multicolumn{1}{c|}{(\%)} &
\colhead{(\%)} &
\colhead{(\%)}
}
\startdata
NGC0628 &   52 &   48 &   34 &   66 &   43  &  57  &  18  &  82 \\
NGC3351 &   44 &   56 &   77 &   23 &   21  &  79  &  24  &  76 \\
NGC3627 &   48 &   52 &   29 &   71 &   31  &  69  &  12  &  88 \\
NGC4254 &   58 &   42 &   23 &   77 &   45  &  55  &  11  &  89 \\
NGC4321 &   57 &   43 &   46 &   54 &   38  &  62  &  24  &  76 \\
NGC4535 &   60 &   40 &   43 &   57 &   45  &  55  &  22  &  78 \\
NGC5068 &   16 &   84 &   88 &   12 &   14  &  86  &  73  &  27 \\
NGC5194 &   59 &   41 &   14 &   86 &   43  &  57  &  7  &  93 \\
\hline
Median  &  57  &  48  &  43  &  66  &  43  &  62  &  22  &  82  \\
Mean  &  49  &  51  &  44  &  56  &  35  &  65  &  24  &  76  \\
\enddata
\tablecomments{Fraction of CO and \hii-associated \ha\ emission in the
  overlap and non-overlap regions of the maps at 140\,pc
  resolution. The four left columns list the fraction of sightlines
  (i.e. by number). The remaining columns list the fraction of the
  flux in each tracer). 
  {\bf Note that the medians are not re-normalized, thus the sum will not add up to 100\%.}
}
\end{deluxetable*}

\textit{Relation to morphology --} In most galaxies, the CO emitting
sightlines form coherent, kiloparsec-scale structures such as spiral
arms, bars, or rings in the disk. The sightlines probing \ha\ emission
appear more irregular and patchy than the CO emission, though they
generally trace out the same galactic structures as the CO. {\it Overlap}
sightlines with both CO and \ha\ emission appear prevalent in the centers of
almost all targets (indicated by the yellow ellipse in
Fig.\,\ref{fig:allgals_olap_gallery_140pc}) and along the major
galactic structures traced by CO emission.

In several galaxies with strong spiral arms, we find a pronounced
offset between sightlines probing \ha\ emission and those with {\em CO
  only}. This is most evident along the spiral arms of NGC\,0628,
NGC\,4321, and NGC\,5194, but the same pattern can be seen in
NGC\,4254, and NGC\,4535. The sense of the
offset is that the sightlines with \ha\ ({\em both overlap and
  \ha\ only}) tend to lie along the convex side of the arm, offset
from the {\em CO only} ones. Assuming that the spiral arms are trailing, 
this implies that the {\em \ha\ only} and {\em overlap} sightlines appear 
offset downstream from the {\em CO only} sightlines (when observing them inside their co-rotation).

The overlap in the central regions appears strong for the barred
galaxies NGC\,3351, NGC\,3627, NGC\,4321, and NGC\,4535. In several of
these targets, we see {\em overlapping} CO and \ha\ in the central region,
and then again in the region outside the stellar bar. Along the bar itself, we
see predominantly {\em CO only} emission, though this emission is not always present
along the full length of the bar (e.g., see NGC\,3351).
A similar feature appears in NGC\,5194, with {\em CO only} sightlines appearing along segments of the inner spiral arms
\citep[features associated with strong streaming motions by][]{meidt13}.

Here and throughout our analysis, NGC\,5068 shows a distinct
morphology: a patchy, incoherent distribution of CO and \ha\ emitting
regions. This galaxy has the lowest stellar mass, lowest metallicity
and latest Hubble type in our sample. It resembles M33, NGC\,300 and the
Large Magellanic Cloud, where similar observations have found tracers
of gas and star formation to be incoherent on small scales
\citep{kawamura09,schruba10,onodera10,gratier12,faesi14}. Those studies have
postulated that star formation in dwarf spirals proceeds in a
stochastic way, without any large-scale organization.

\begin{figure*}[tbh]
\includegraphics[width=\textwidth,angle=0]{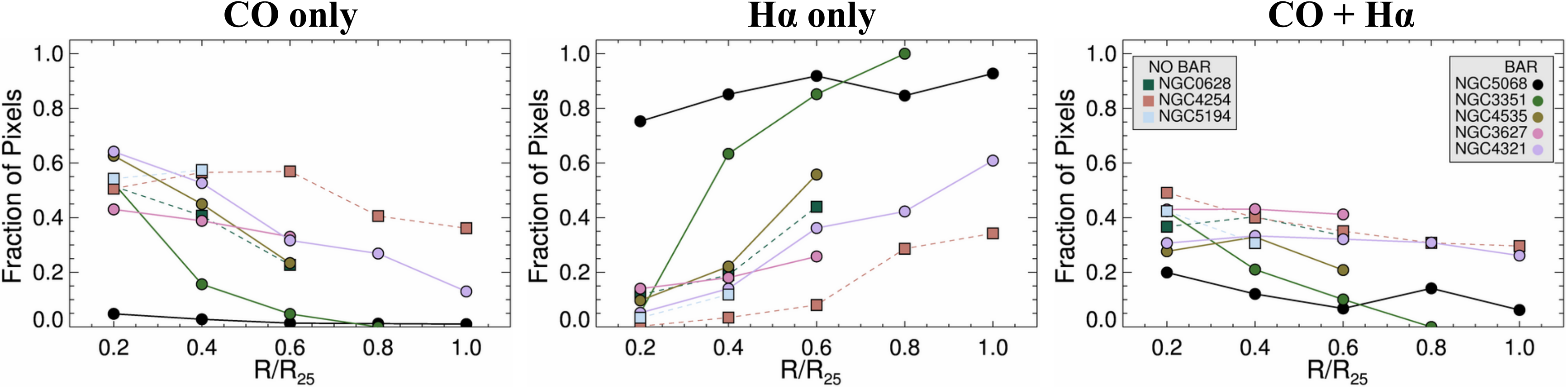}\hfill
\caption{Variation of the overlap fraction at 140\,pc resolution as a
  function of galactocentric radius. The $x$-axis is normalized by
  $R_{25}$, and the curves are color-coded by increasing stellar
  mass. Barred galaxies are shown with circles and solid lines, non-barred galaxies   
  with squares and dashed curves. The classification of Tab.\,\ref{tab:sample} is adopted. 
\label{fig:olap_rgal_140pc}
}
\end{figure*}

\textit{Trend with Galactocentric Radius --} Visual inspection of the
maps in Figure~\ref{fig:allgals_olap_gallery_140pc} suggests that many
of the {\it \ha\ only} sightlines occur towards the outer edges of our
maps.  We quantify this in Figure~\ref{fig:olap_rgal_140pc}, where we
plot the {\it CO only}, {\em \ha\ only} and {\it overlap} fractions in
annuli of width $0.2R_{25}$. Indeed in all galaxies except NGC\,5068,
there is a clear trend for the {\em \ha\ only} fraction to increase
with increasing galactocentric radius. This increase seems more
pronounced in barred galaxies (plotted with circles and solid lines in
Figure~\ref{fig:olap_rgal_140pc}). At the same time the
radial distribution of the {\em CO} sightlines (both {\em CO only} and
{\em overlap}) shows on average a stronger (decreasing) trend for
barred galaxies than for the three non-barred ones, however, also with
a much wider spread in values. 
A possible explanation for these trends is the fact that photoionization bubbles expand more quickly in low-density media because the Str\"omgren radius scales as $\rho^{-2/3}$, thus larger HII regions are expected at larger radii where the gas surface density is presumably lower. Conversely, this decrease in gas surface density should lead to more isolated patches of molecular gas. As many other galaxy parameters vary with galactocentric radius, e.g., metallicity,
a secure interpretation of these trends requires further investigation using a larger sample with better sampling of galaxy morphological type and stellar mass.

\subsection{Overlap between CO and \ha\ as function of spatial scale}
\label{subsec:2Dmaps_scale}

\begin{figure*}[tbh]
\includegraphics[width=\textwidth,angle=0]{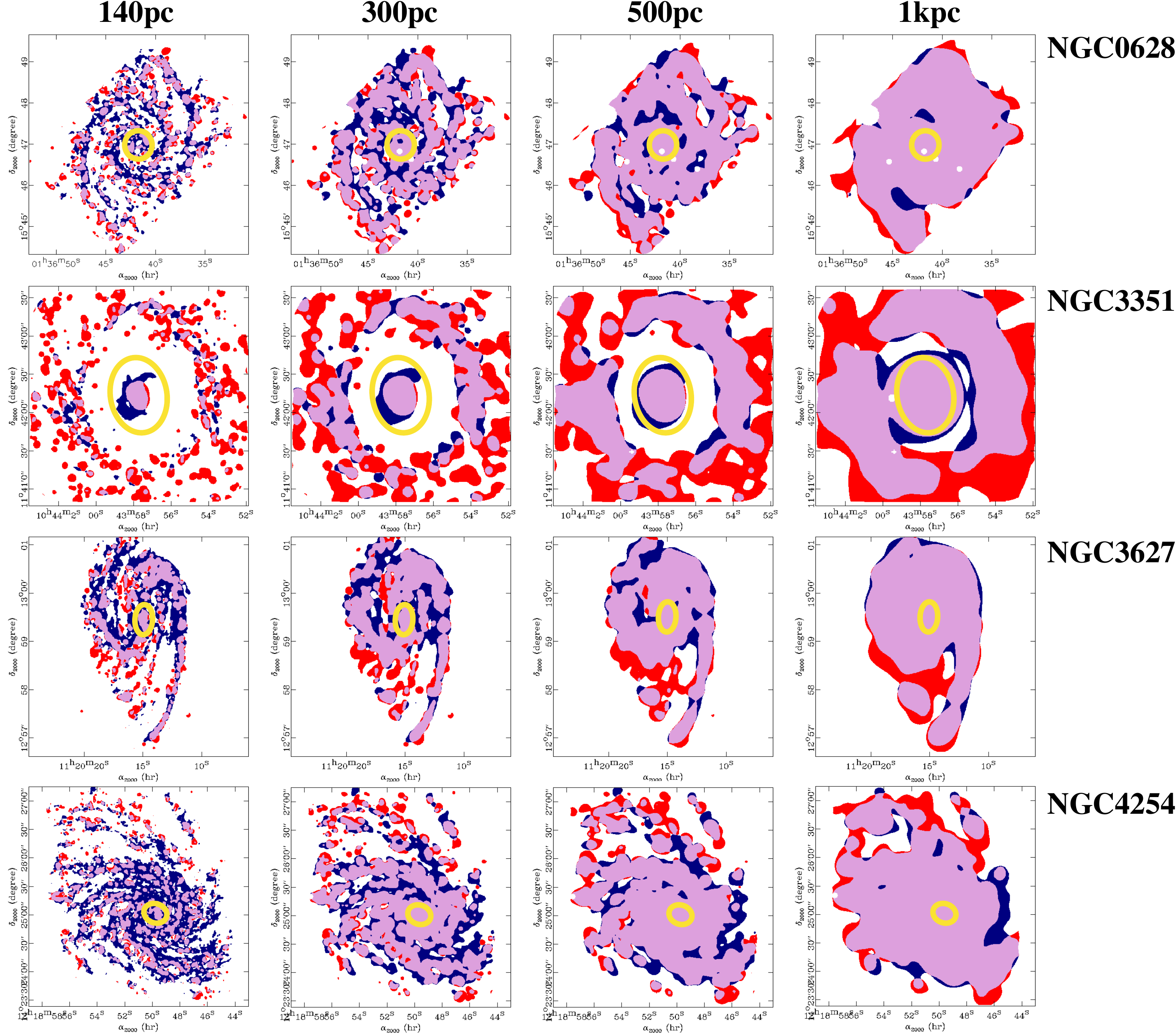}\hfill
\caption{Galaxy maps showing regions with overlapping CO and
  \ha\ emission (lavender), {\em CO only} (dark blue) and {\em
    \ha\ only} emission CO (red) at spatial resolutions of 140, 300
  and 500 and 1000\,pc ({\em left to right}) for NGC\,628, NGC\,3351,
  NGC\,3627 and NGC\,4254.({\em top to bottom}).  
\label{fig:allgals_olap_gallery_1}}
\end{figure*}

\begin{figure*}[tbh]
\includegraphics[width=\textwidth,angle=0]{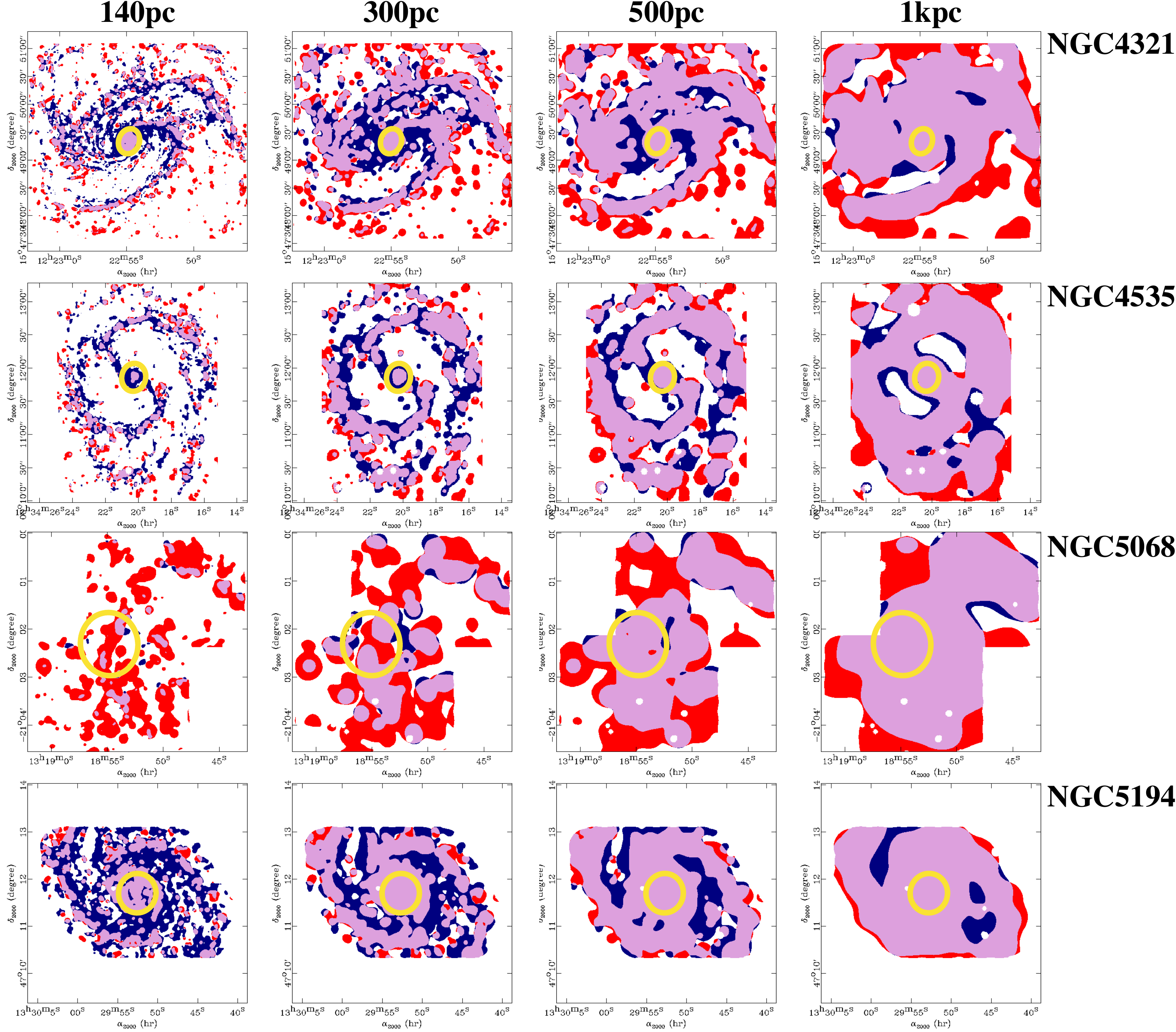}\hfill
\caption{Galaxy maps for NGC\,4321, NGC\,4535, NGC\,5068
  and NGC\,5194 ({\em top to bottom}). (For details see caption of Fig.\,\ref{fig:allgals_olap_gallery_1}.)
\label{fig:allgals_olap_gallery_2}}
\end{figure*}


\begin{figure*}[tbh]
\centering
\includegraphics[width=\textwidth,angle=0]{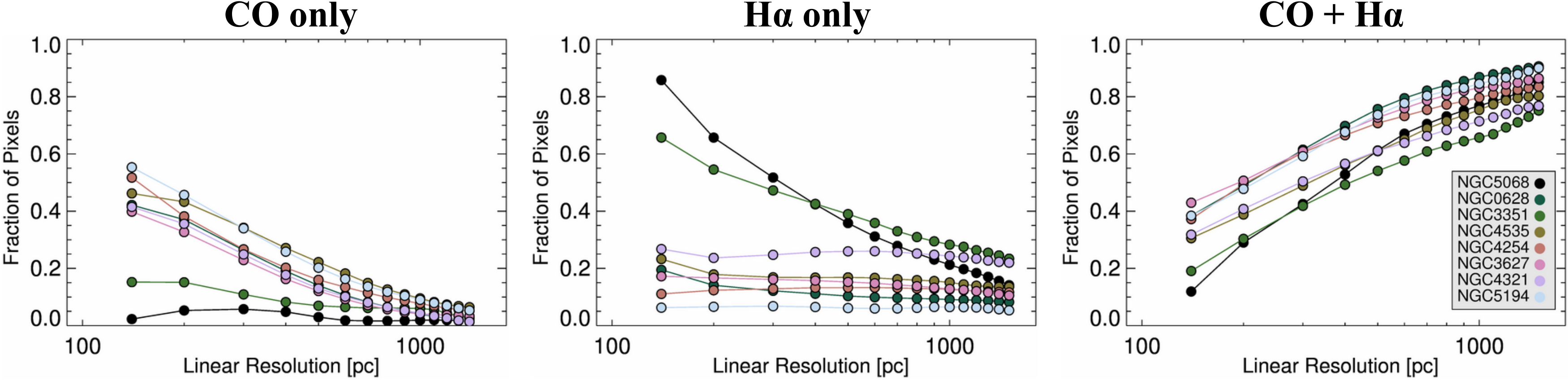}\hfill
\caption{Fraction of pixels belonging to the different categories as a
  function of spatial scale. From left to right, the three panels
  show the variation of {\em CO only} regions, {\em \ha\ only} regions and regions with {\em overlapping} CO and
  \ha\ emission. As in the previous figures the pixel fractions are defined using the total number of pixels containing emission from either CO and/or \ha\ (see text for details).
\label{fig:ofrac_summary_plot}}
\end{figure*}

\begin{figure*}[tbh]
\includegraphics[width=\textwidth,angle=0]{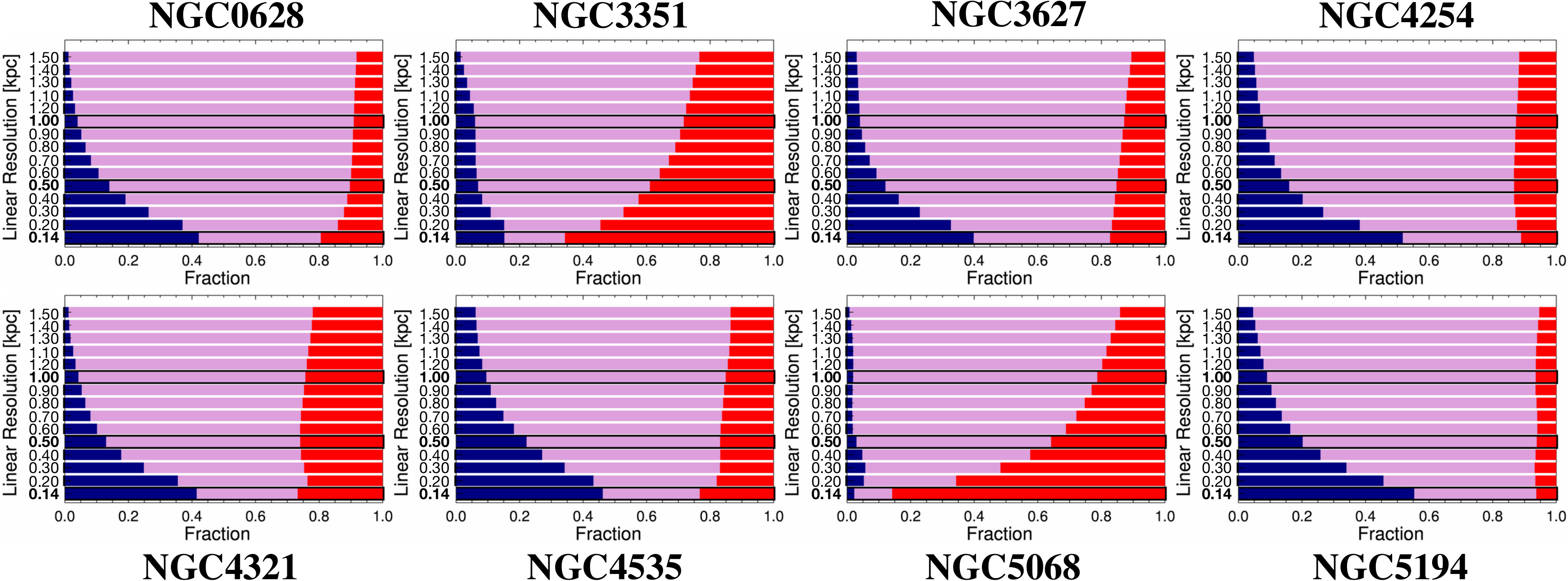}\hfill
\caption{Bar graphs indicating the fraction of sightlines with CO and
  \ha\ emission across the range of spatial scales that we explore. The order of the galaxies and the color code are the
  same as for Fig.\,\ref{fig:piechart_gallery_p1}. The spatial scales
  of 140pc, 500pc and 1kpc are highlighted by a black outline.
\label{fig:barchart_gallery_p1}}
\end{figure*}

Figs.\,\ref{fig:allgals_olap_gallery_1} and~\ref{fig:allgals_olap_gallery_2} illustrate the spatial
distribution of CO and \ha\ emission as we vary the resolution of the
maps (showing maps at 140\,pc, 300\,pc, 500\,pc, and 1000\,pc). In Figs.\,\ref{fig:ofrac_summary_plot} and
\ref{fig:barchart_gallery_p1}, we plot the covering fractions of {\em
  CO only}, {\em \ha\ only}, and {\em overlap} regions as a function
of spatial scale (from 140\,pc up to 1.5\,kpc in steps of 100\,pc starting at 200\,pc).

\textit{From 140\,pc to 500\,pc resolution --} As we blur the maps
from 140\,pc to 500\,pc resolution, sightlines where CO and
\ha\ emission coincide become more prevalent in all galaxies. By
500\,pc resolution, {\em overlapping} (lavender) sightlines dominate
($>$60\%) while {\em CO only} sightlines make up $<$25\% of the
emitting sightlines in all targets. For most targets, we see a significant
slow-down in the increase of the fraction of {\em CO only} sightlines at spatial scales of
$\sim$150-200\,pc. The exceptions are NGC\,4254 and NGC\,5194, where
the {\em CO only} fraction continuously decreases from 140\,pc to
larger scales.

Most of the evolution of {\em overlap} fraction and structure of {\em
  CO only} emission occurs over this range, $140{-}500$~pc. As the
resolution degrades, the detailed morphology visible at high
resolution evolves toward a simple, amorphous geometry. Structures
that were prominent at sharper resolution, such as spiral arms and
bars, disappear and emission fills in the region between galactic
structures. As this happens, the region covered by {\em overlapping}
CO and \ha\ emission grows from patches to form coherent structures
along spiral arms and stellar bars. Over this range, in all but two
galaxies (NGC\,3351 and NGC\,5068) the {\em \ha\ only} covering
fraction shows less significant evolution than the {\em CO only} or
{\em overlap} fraction. Three-quarters of our sample show no variation
in {\em \ha\ only} covering fraction for scales of 300\,pc and
larger. In these cases, the {\em \ha\ only} regions often appear as
mostly isolated, compact regions, often located at large
galactocentric radius.

\textit{Beyond 500\,pc resolution --} As we blur the maps to
resolutions greater than 500\,pc, the {\em overlap} sightlines tend to
become unified structures that include the galaxy nucleus. By 1.0\,kpc
resolution, the {\em CO only} (dark blue) sightlines tend to no longer
trace galactic structures, but rather appear spread across the
galaxy. This heavy overlap between CO and \ha\ emission at coarse resolution
agrees with numerous previous studies showing a tight correlation
between these quantities at this resolution
\citep[e.g.,][]{bigiel08,leroy08,liu11,leroy13,bolatto17}.

Overall, the changes in morphology and covering fraction beyond
500\,pc resolution appear less pronounced than those between 140\,pc
and 500\,pc. For the majority of our sample, the most significant
changes in the fraction of {\em overlapping} (lavender) sightlines
occur between $140$ and $500$\,pc resolution. The fact that this
overlap emerges only after averaging agrees with previous observations
of increasing scatter in the correlation between molecular gas and
star formation tracers with decreasing spatial scale
\citep[e.g.,][]{bigiel08,schruba10,onodera10,leroy13,kruijssen19}.


\section{Discussion}
\label{sec:discussion}

We have analyzed a sample of matched resolution, high completeness,
wide area CO and \ha\ maps, which trace molecular gas and high-mass star
formation. Consistent with previous observational studies targeting individual
galaxies \citep[e.g. M33 and M51,][]{schruba10,meidt13}, we find a large reservoir of molecular gas not directly
associated with high-mass star formation at high spatial
resolution. A significant non-star-forming molecular gas reservoir in galaxies is expected from numerical simulations \citep[e.g.][]{semenov17}. This component has been almost impossible
to identify in previous, low resolution surveys because the apparent overlap between tracers of molecular gas and star formation increases with degrading
resolution. At the coarsest resolutions that we study, $\sim 1.0-1.5$~kpc,
CO and \ha\ emission appear almost completely coincident. 

We also clearly observe the effects of galactic dynamics in the maps:
bars, spiral arms, and central regions all stand out in our analysis.
This dynamically-induced complexity indicates that the timescales associated with different steps in the star formation process (e.g. cloud formation, cloud dispersal) may depend on galactic environment. We thus expect timescale estimates based on galaxy-wide measurements and simple cycling arguments to mask intrinsic variation of star formation timescales within galaxies. 

With this caveat in mind, we derive rough estimates for 
the time required for the molecular gas in our sample galaxies to cycle between a quiescent and actively star-forming state (see
\S \ref{subsec:temporal}). Our measurements imply that this timescale is $\sim1$ to 2 times longer than the characteristic lifetime of an {\sc Hii} region. This is broadly consistent with estimates for molecular cloud lifetimes, but much shorter than the global molecular gas depletion times of $\sim2$\,Gyr observed for nearby galaxies \citep[]{leroy08}. An investigation of molecular cloud lifetimes across a subset of PHANGS galaxies using an alternative formalism is presented by \citet{chevance19}.

\subsection{Quiescent molecular gas at high resolution, overlap at low resolution}

The relative contributions of {\em CO only}, {\em \ha\ only} and {\em  overlap} sightlines vary dramatically with spatial scale. At $\sim 140$~pc resolution, we find a roughly equal number of CO-emitting sightlines (at and above our $\rm H_2$ surface density threshold of $\rm 12.6\,M_{\odot}pc^{-2}$)  with and without associated \ha\ emission  in our sample galaxies. The {\em CO only} sightlines contribute $\sim40$\% of the total CO flux above this threshold. This implies the presence of a significant reservoir of quiescent molecular gas. 

This apparently quiescent molecular gas could have several distinct physical origins. This gas may be genuinely non-star-forming, due to large-scale dynamical effects that stabilize the gas  \citep[e.g.,][]{meidt13,meidt19} or the disruptive effect of stellar feedback \citep[e.g.][]{krumholz05,ostriker10}. Alternatively, it could be forming stars that are not massive enough to generate the ionizing photons needed for significant \ha\ emission. High-mass stars could be forming behind a screen of dust that obscures \ha\ photons (see Table~\ref{tab:coha_data}). Or the gas could still be in the process of collapsing, reflecting recently formed molecular clouds that have not yet ``ramped up'' star formation \citep[e.g.,][]{lee16}. 

\textbf{Quiescent gas and spatial scale:} This quiescent molecular gas is only accessible thanks to the high physical resolution of our data. For our adopted CO and \ha\ thresholds, the fraction of {\em CO only} sightlines decreases from $\sim40$\% at 140\,pc resolution to $\lesssim15$\% at 500\,pc. Our adopted fiducial scale of 140\,pc is a pragmatic choice, not a physical one. For five galaxies in our sample, the best matching spatial resolution of our data is 90\,pc or better, allowing us to test whether the fraction of {\em CO only} sightlines in galaxies is larger at higher physical resolution. We repeated the analysis described in Section~\ref{sec:method} for the CO and \ha\ maps of these galaxies at 90\,pc resolution, using a threshold of $\rm 19.5\,M_{\odot}pc^{-2}$ (i.e. the 3$\sigma$ mass surface density sensitivity of our NGC\,5194 data at 90\,pc resolution). 
We find that the fraction of {\em CO only} sightlines in all galaxies shows a relative increase of 5 to 20\% as the resolution improves from 140 to 90\,pc. The fraction of {\em \ha\ only} sightlines typically shows a similar relative increase, while the fraction of {\em overlap} sightlines shows a relative decrease of $\sim30$\% as the resolution improves. 

At coarser resolution than 140~pc, the quiescent CO emission rapidly vanishes as the resolution is degraded. The rapid decrease in {\em CO only} sightlines between 140 and 500\,pc resolution is accompanied by a significant increase in the fraction of {\em overlap} sightlines (from $\sim30$\% to  $\sim65$\%) across this range of scales.

For most of our sample, the visible morphology of CO emission also changes over this spatial range: as the resolution degrades, the initially sparse, patchy {\em overlap} sightlines become more common, while the {\em overlap} regions spread out, replacing the {\em CO only} regions and covering the main morphological features of the galaxy. Scale-dependent variations in the fraction and spatial distribution of the {\em \ha\ only} sightlines are less marked in our current, small sample of galaxies. This is because the  {\em \ha\ only} regions tend to be relatively isolated, distinct regions towards the outer parts of our maps.

At resolution coarser than $\sim500$\,pc, the {\em overlap} sightlines cover most of the mapped area in all targets. At these scales, differences in the morphology of the CO and \ha\ emission disappear. Even major morphological features of galaxies (e.g., prominent spiral arms, very luminous \hii\ regions in outer disks)  start to ``wash out'' and are mostly indistinguishable on  spatial scales above $\gtrsim$1.0\,kpc.

Our results agree with previous observations showing how 
star formation scaling relations -- i.e. the power law scaling between the molecular gas and star formation rate (SFR) surface density, where SFR is typically traced by \ha\ emission -- vary with spatial scale. Several studies have demonstrated that the scatter in the molecular gas depletion time decreases as the spatial scale of the molecular gas and SFR surface density measurements increases. These works include high resolution studies of individual galaxies \citep[e.g.][]{blanc09,schruba10,onodera10,kreckel18,kruijssen19} and coarser resolution studies of larger galaxy samples \citep{bigiel08,leroy13}. To our knowledge, our work here represents the best combination of resolution ($140$~pc across the sample) and sample size ($8$ galaxies) to date.


\subsection{Temporal interpretations}
\label{subsec:temporal}

\begin{deluxetable}{l|cccc}
\centering
\tablecaption{Inferred Relative Gas Cycling Timescales by Sightlines at Various Scales.\label{tab:tcloud_bynum}}
\tablehead{
\colhead{Name} &
\multicolumn{4}{c}{$ f_{scale} = \frac{f_{CO-only}+f_{overlap}}{f_{\ha-only} + f_{overlap}}$} \\
 & 
\colhead{140\,pc} &
\colhead{300\,pc} &
\colhead{500\,pc} &
\colhead{1000\,pc} 
}
\startdata
NGC\,0628   &   1.4   &   1.2  &   1.0    & 0.9 \\
NGC\,3351   &   0.4   &   0.6  &   0.7    & 0.8 \\
NGC\,3627   &   1.4   &   1.1  &   1.0    & 0.9 \\
NGC\,4254   &   1.8   &   1.2  &   1.0    & 0.9 \\
NGC\,4321   &   1.3   &   1.0  &   0.9    & 0.8 \\
NGC\,4535   &   1.4   &   1.3  &   1.1    & 0.9 \\
NGC\,5068   &   0.1   &   0.5  &   0.7    & 0.8 \\
NGC\,5194   &   2.1   &   1.4  &   1.2    & 1.0 \\
\hline 
Median  &  1.4  &   1.2 & 1.0 & 0.9\\
Mean  &  1.2    &  1.0 & 1.0 & 0.9\\
\enddata
\tablecomments{{\bf The reported values are derived using Eq.\, \ref{eq:timescale} and
need to be multiplied by the assumed $t_{H\alpha}$, i.e. the time an {\sc Hii} region is visible above our thresholds.} See text for discussion of reasonable values for $t_{H\alpha}$.
}
\end{deluxetable}

Under the assumption of simple steady-state cycling, every star-forming unit progresses from a cold gas cloud that is not yet forming stars ({\em CO only}) to a cold gas cloud that is actively forming massive stars ({\em overlap} regions) to a region of massive young stars from which the cold gas has been entirely dispersed ({\em \ha\ only}). In this case, the covering fractions that we measure are a reflection of the time spent by a star-forming region in each of these different evolutionary phases. 

Versions of this steady-state cycling heuristic have been used by \citet[][]{kawamura09} to estimate the lifetimes of molecular clouds in the LMC, and by \citet[][]{gratier12} and \citet{corbelli17} to characterize the cloud life-cycle in M\,33. Recently, \citet[][]{battersby17} applied a similar approach to Milky Way observations to determine the timescales for dense molecular gas structures (i.e. clumps) to evolve from starless to
star-forming. \citet{schruba10} invoked a similar cycling scenario to explain the scatter and bias in the molecular gas depletion time
measured at high spatial resolution in M\,33. \citet{feldmann11} also considered this phenomenon in terms of the scatter \citep[see
  also][]{leroy13}. Building on this work, \citet{kruijssen14} and
\citet{kruijssen18} developed a general formalism to derive the evolutionary timescales of star-forming regions from concentric apertures centered on peaks identified at different wavelengths. This formalism has been applied to CO and \ha\ maps of the spiral galaxy NGC\,300, an M\,33 analog, in order to estimate the typical lifetime of molecular clouds and characteristic duration of star formation feedback in that galaxy  \citep{kruijssen19}. Analogous arguments have been used to infer timescales in other fields, including, e.g., to measure the life cycle of protostars \citep[e.g., see review in][]{dunham14}. All these methods require an externally calibrated fiducial clock, e.g., the time $t_{H\alpha}$ over which an H{\sc ii} region is visible, to translate between population/mass/area statistics and an absolute timescale. 

Here we apply a simple version of this approach to molecular gas that is visible above a fixed surface density threshold, treating individual pixels as discrete star-forming units. The lifetime of the cold molecular gas in a star-forming unit is then

\begin{equation}
\label{eq:timescale}
t_{gas} = t_{H\alpha } \times \frac{f_{CO-only}+f_{overlap}}{f_{\ha-only} + f_{overlap}} = t_{H\alpha } \times f_{scale}~,
\end{equation}

\noindent 
where $f_{CO}$ is the fraction of {\em CO only} sightlines, $f_{H\alpha}$ the fraction of {\em \ha\ only} sightlines, and $f_{overlap}$ the fraction of {\em overlap} sightlines. Then $f_{scale}$ represents the scaling factor to translate from the fiducial timescale, here $t_{H\alpha }$, to the lifetime for a cold gas structure.

\subsubsection{Timescales}
\label{subsec:cycling_results}

Table \ref{tab:tcloud_bynum} reports the results of Eq. \ref{eq:timescale} applied to our sample at various scales using the numbers derived in \S\,\ref{sec:results}. As the resolution improves (i.e., the spatial scale decreases), regions with {\em overlapping} \ha\ and CO emission tend to occupy a diminishing fraction of sightlines (see Fig.\,\ref{fig:barchart_gallery_p1}). The ratio $f_{scale}$ thus tends to increase for galaxies where {\em CO only} sightlines are prominent, while it decreases rapidly for galaxies where {\em \ha\ only} sightlines dominate (i.e., NGC\,3351 and NGC\,5068). If we were to improve upon our 140~pc resolution, we would expect these trends to continue. This highlights some natural next steps to improve on the current data treatment (see \S\,\ref{subsec:caveats}).

For most of our targets, these measurements imply a median CO visibility timescale of $1.4\times$ the \ha\ visibility timescale $t_{H\alpha}$ at 140\,pc resolution. Typical estimates for the visibility time of \ha\ are $5{-}10$~Myr\footnote{Based on population synthesis models, \citet{kennicutt12} estimate that 90\% of emission is emitted within $\sim 10$~Myr with a mean age of $\sim 3$~Myr for the associated stellar population \citep[for other estimates, see, e.g.][]{leroy12,haydon18}.}. Binaries, extinction, and other effects may complicate the picture, but taking these numbers at face value, Table\,\ref{tab:tcloud_bynum} suggests molecular cloud lifetimes of $\sim 10{-}15$~Myr at our best resolution of 140\,pc. 

While there is a clear dependence on scale (and the data have some limitations, see \S\,\ref{subsec:caveats}), these short timescales are in reasonable agreement with measurements of molecular cloud lifetimes in other galaxies in the literature (see above). For the specific galaxies in our sample besides NGC\,5068 and NGC\,3351, our estimates also agree within a factor of $\sim2$ with cloud timescale estimates obtained by \citet{chevance19} applying the formalism of \citet{kruijssen18} to these systems. This robustness to adopted methodology suggests that both sets of results do indeed reflect the true underlying cycling times.

The order of magnitude of $t_{gas}$ also agrees well with theoretical estimates of the timescale for star formation. It is comparable to the crossing or free-fall time for individual molecular clouds \citep[e.g., see][for typical cloud properties]{heyer09,fukui10}, including those in our sample \citep[e.g., see][for direct estimates]{utomo18}. Many models have taken either the crossing time \citep[e.g.,][]{elmegreen00} or the free-fall time \citep[e.g.,][]{krumholz05} as the relevant timescale for star formation. Numerical models find similar results, e.g., based on recent highly resolved simulations of galaxy disks, \citet{semenov18} quote star formation timescales of $\sim$5-15\,Myr. Our measurements support the idea that the onset of star formation and dispersal of clouds occurs over roughly this timescale.

\subsubsection{A more dynamic view}
\label{subsec:cycling_dynamics}

The simple evolutionary sketch underlying Equation \ref{eq:timescale} may be too simple to capture the evolution of real regions in the molecular ISM.
Recent numerical simulations of galaxies suggest that molecular gas may cycle in and out of a bound, cloud-like state several times before finally participating in the formation of stars \citep[e.g.][]{dobbs15,semenov17}, and that this may explain the discrepancy between the molecular gas depletion times measured for entire galaxies \citep[$\sim2$\,Gyr, e.g.][]{leroy08} and individual molecular clouds \citep[a few hundred Myr, e.g.][]{evans09,heiderman10}. In this case, not every patch of visible CO emission will form stars in the near future. 

Moreover, grouping together CO-bright stellar bars and \ha -bright regions at large galactocentric radius will not capture local, physical cycling between gas and star formation. Even in star-forming regions, the visible CO-\ha\ offsets associated with spiral arms \citep[e.g., prominently observed in M\,51, see e.g. ][]{schinnerer13,schinnerer17} suggest that a one-dimensional flow model \citep[e.g.,][]{meidt15,dobbs13} may be more appropriate than stochastic cycling.

The trends of decreasing $f_{scale}$ with increasing resolution for 6 of our 8 galaxies might also reflect this more dynamic view. For example, in the \cite{semenov17} model, gas cycles in and out of a star-forming state several times before a fraction of the star-forming gas is converted into stars. Thus with increasing resolution we start to probe gas cycling between a star-forming and non-star-forming state which is setting the timescale $t_{gas}$ we infer. The inverse trend in $f_{scale}$ observed for especially NGC\,5068 and potentially NGC\,3351 may be set by their rather short timescale for {\em localized} star formation to occur. This timescale is set by stellar feedback that quickly removes gas from the star-forming state \citep[see also, e.g.,][]{kruijssen19} and such galaxies are in the ``self-regulation'' regime.

\subsection{Visible effects of galactic dynamics}

Our maps show clear signatures of galactic dynamics. In addition to the high overlap fraction in most galaxy nuclei, we observe signatures of the suppression of star formation in bars and offsets between \ha\ and molecular gas emission along spiral arms.

\smallskip
\textbf{Central regions:} All our eight galaxies show prominent {\em overlap} regions in the central 1\,kpc (see Fig.\,\ref{fig:allgals_olap_gallery_1}).
Close inspection of the CO intensity maps and \ha\ images presented in Fig.\,\ref{fig:gallery_p1} suggests that the overlap is arising mainly due to
\ha\ emission related to 
(a) central star formation mostly distributed in so-called nuclear star-forming rings \citep[e.g. NGC\,3351 and NGC\,4321 are listed in the AINUR catalog of ][]{comeron10} or along the stellar bar (in NGC\,5068) or 
(b) the presence of AGN (e.g., as good examples see NGC\,3627, NGC\,5194) where
the \ha\ emission is coming from the ionized gas in the narrow line region \citep[e.g. in NGC\,5194;][]{blanc09}. Thus, unlike the Central Molecular Zone (CMZ) in our own Galaxy, 
other nearby galaxies can show significant star formation, especially in the
nuclear star-forming rings that are typically found in barred galaxies \citep[see also review by ][]{kennicutt98b}. Our inability to discriminate between \ha\ emission arising from truly star-forming structures or from a central
AGN is a caveat for the interpretation of our results in galaxy centers (see \S\,\ref{subsec:centers} for further discussion).

\smallskip
\textbf{Stellar bars:} Observational and theoretical work on the role
of stellar bars in star formation still draws somewhat conflicting
conclusions about the roles for these structures. It is clear that bars help drive material to the central regions of galaxies, but whether they enhance or suppress star
formation along the way remains debated \citep[e.g.,
  see][]{meidt13,renaud16,james18}. Two of our targets, NGC\,3351 and
NGC\,4535, show strong signatures of suppressed star formation along a
strong stellar bar. Two more, NGC\,3627 and NGC\,4321 show some
evidence of this phenomenon.

The signatures are a clear gap between the
spiral arms, where overlapping CO and \ha\ emission appears common,
and the center. CO emission remains visible in this gap, while \ha\ emission appears to be largely absent. In NGC\,3627 particular, the phenomenon is harder to identify
because of the brightness of star-forming complexes at the bar ends 
\citep[e.g., see][]{beuther17}. A similar depression in star formation has been identified along the spiral arms
in NGC\,5194 and attributed to streaming
motions along the spiral arms (not a stellar bar) by
\citet{meidt13}. 

In theoretical calculations, the high shear and/or diverging streamlines (caused by changing orbits) present at the location of the gas lanes along the stellar bar either prevent clouds from forming or tears them apart before star formation can happen \citep[e.g.][]{athanassoula92,regan97}. Observationally, this ``star formation desert'' (along the stellar bar) is not evident in
all barred galaxies \citep[see e.g.][]{sheth00,james18}. This suggests that the exact bar properties and/or the time evolution of bars may also be important to the suppression of star formation. 

Overall, our data provide support for the idea that star formation in
molecular gas is inhibited along many stellar bars. Despite the presence
of bright CO emission, {\sc Hii} regions appear less common in these
regions.

\smallskip
\textbf{Spiral arms:} For several galaxies we find a pronounced spatial
offset between CO and \ha\ emission along
spiral arms. In these cases, we observe a general pattern where {\em
  CO only} sightlines are more prominent on the concave side, while
{\em overlap} sightlines are found along the convex side of a spiral
arm, with {\em \ha\ only} sightlines sitting outside the {\em overlap}
regions away from the {\em CO only} regions.

Figure \ref{fig:allgals_olap_gallery_140pc} shows impressive examples
of this phenomenon in the grand design spiral galaxies NGC\,0628
\citep[particular at smaller galactocentric radii; see also Fig. 1 in
][]{kreckel18} and NGC\,5194
\citep[e.g.][]{schinnerer13,schinnerer17}. The spiral arms emanating
from the stellar bars in NGC\,4321 and NGC\,4535 also
exhibit offsets along significant arm segments. This suggests that a
similar mechanism causes this offset in grand-design spirals and
spiral arms dynamically linked to stellar bars. 
Assuming the spiral arms are trailing, the \ha\ offsets occur on the downstream side consistent with expectations for a rotating spiral density wave (inside co-rotation) \citep[for a sketch see, e.g., Fig.\,1 of ][]{pour-imani16}. 

\smallskip
\textbf{Trends with global properties:} We compared
the overlap fraction and amount of apparently quiescent CO emission to
the integrated properties of the host galaxy, including stellar mass ($\rm M_{star}$), star
formation rate (SFR) and specific SFR (sSFR$\equiv$SFR/M$_{star}$). While our sample of $8$
galaxies represents a significant improvement compared to previous
work, a much larger sample will be required to measure quantitative
relationships. Still, some trends are already evident from these first results.

First, targets that have a large fraction of {\it CO only} sightlines at 140\,pc resolution are all
more massive than $\rm log(M_{\star}[M_{\odot}])\,=\,10.2$, tend to have higher SFRs ($\rm SFR[M_{\odot}yr^{-1}]\,>\,2$), and also tend to lie above the average stellar mass--SFR relation. Such galaxies tend to be rich in molecular gas compared to lower mass and more quiescent galaxies. This result suggests that, at least at any given moment, a large fraction of the gas in such galaxies may not be immediately associated with high mass star formation. We do caution that our present sample is heavily biased towards such objects.

By contrast, {\it
  \ha\ only} sightlines are most prominent in the two lowest mass and
lowest SFR galaxies which also have the latest (NGC\,5068) and
earliest Hubble type (NGC\,3351) of the sample. Though we do not
include M33 in the analysis, this would appear to be consistent with
the widespread \ha\ and confined CO emission visible in that galaxy
\citep[e.g., see][]{schruba10,druard14,corbelli17}. Here, sensitivity, conversion factor effects, and the choice of thresholds will play an important role, because these are also among our most quiescent targets. 

The one galaxy
where {\it overlap} sightlines dominate (NGC\,3627) exhibits on
average low SFR and sSFR values and is also part of a group currently
undergoing a large-scale interaction.

No trend with the presence or
absence of a (large-scale) stellar bar is evident. Similarly, we find
no clear trend with the presence or absence of an AGN (all galaxies
with $\rm log(M_{\star}(M_{\odot})\,\gtrsim\,10.4$, i.e. 5/8 galaxies, show
evidence for an AGN). This is not surprising given the small sample size, the relatively low number of sightlines directly associated with the galaxy center or bar, and the temporal evolution expected for both phenomena.

\smallskip
\textbf{Does feedback or galactic dynamics drive the CO morphology?} In a series of recent papers, \citet{semenov17,semenov18} put forward a star formation model based on high resolution simulations of gas and star formation in galaxy disks. \cite{semenov18} distinguish two distinct regimes, which may relate to the morphology of CO and \ha\ emission seen in our observations. In the ``self-regulation'' regime, the lifetime of gas in the star-forming state is limited by consumption via star formation and feedback. A stochastic, patchy appearance might be expected, with the CO emission shaped by stellar feedback. By contrast, in the ``dynamics-regulation'' regime, the lifetime of the star-forming gas is limited by dynamical processes. In this case, they expect the star formation to reflect the underlying distribution of the cold gas rather than being shaped by star formation and/or associated feedback. 

In our observations, both the star-forming and quiescent molecular gas trace distinct bars, rings, and arms in most galaxies. In other words, it appears that the star formation traces the underlying molecular gas distribution, which in turn traces the underlying potential (see Fig.\,\ref{fig:olap_rgal_140pc}). The patchy stochastic appearance that we might expect for morphology dominated by self-regulation is most prominent in the outer disk of NGC\,3351 and the low mass galaxy NGC\,5068. This qualitative comparison, as well as the discussion above, demonstrates that the different dynamical environments and host galaxy properties do lead to varying conditions for star formation and that a simple assumption of self-regulation by stellar feedback that results in stochastic, isolated regions is probably too simplistic for many galaxies.


\subsection{Distributions by luminosity vs. distributions by area and role of galaxy centers}
\label{subsec:centers}

For three of our targets, bright CO emission from the
center (see Fig.\,\ref{fig:gallery_p1};
NGC\,3351, NGC\,4321, and NGC\,4535) makes a large fractional contribution to the total CO luminosity
\citep[e.g., see flux-weighted histograms in Figure 1 of ][]{sun18}. 
When we consider the distribution of flux, these
bright centers exert a large influence on the fractions of
emission associated with {\em CO only}, {\em \ha\ only}, and
{\em overlap} regions (see Fig.\,\ref{fig:appendix_summary} and Tab.\,\ref{tab:byflux_bynum}). 
Because they cover relatively little area, the centers
have a weaker effect on the distributions of covering fraction.

Excluding the centers leads to a slightly better agreement between a
flux-weighted approach and a sightline-based approach. There are still
differences between the two approaches, however. In general, the
fraction of luminosity associated with {\em overlapping} CO and \ha\ is
larger than the fraction of {\em overlapping} sightlines. This implies that
the intensity of CO emission is higher in regions that also show
\ha\ emission and, similarly, that the brighter \ha\ emission is associated
with regions of higher CO emission.

Brighter CO emission in regions with \ha\ emission could be expected
if star formation increases the temperature and excitation of the CO
gas. This appears to be the case in a subset of our targets (T. Saito; priv. comm.). It may also reflect that the molecular gas structure
differs between actively star-forming regions and quiescent
regions. This also appears to be the case, with higher surface density
gas being associated with active regions. 

\subsection{Impact of methodology and next steps}
\label{subsec:caveats}

We adopt relatively simple methods to identify pixels tracing CO emission and likely {\sc Hii} regions that could bias our results. It also highlights natural next steps as we apply the method to a larger sample of galaxies.

Our fixed CO threshold leads to variable completeness across our sample \citep[see][]{sun18}. Specifically, using our fixed threshold, we recover a lower fraction of the total CO emission in low mass galaxies and outer disks compared to massive spiral galaxies. \cite{hughes13} showed that molecular clouds in such systems have lower molecular gas surface densities \citep[see also][]{sun18, schruba19}. This appears strongest in NGC~5068 and the outer part of NGC~3351, but may be present at large radii in most of our sample. 

A bias against recovering all of the CO emission present affects the inferred $f_{scale}$, driving the apparent cycling time to artificially low values. Future work will need to address that the CO emission in some regions of our sample is faint compared to our sensitivity. This can be achieved by using statistical methods that account for the effects of noise and by informing the measurements using the known total CO flux from each region of the galaxy.

Local variations of the CO-to-H$_2$ conversion factor, $\rm \alpha_{CO}$, could also impact our analysis. In this case, our estimates for CO-emitting fractions ``by flux'' would no longer straightforwardly correspond to a mass fraction of the molecular gas reservoir. Moreover, the meaning of our adopted threshold would change across the galaxy. Previous work does not suggest enormous $\alpha_{\rm CO}$ variations across this specific sample. \citet[][]{sandstrom13} derived $\rm \alpha_{CO}$ at $\sim$kpc resolution for five galaxies in our sample (NGC\,0628, NGC\,3351, NGC\,3627, NGC\,4254, NGC\,4321) and saw no significant variations across the disks or with radius (excluding some central depression which cover only a small area in our analysis). Similarly, \citet[][]{leroy17a} found a basically constant $\rm \alpha_{CO}$ across the PAWS area in NGC\,5194. Nonetheless, in a preliminary comparison, we do find some modest differences between molecular cloud properties and CO excitation in the {\em CO  only} regions and {\em overlap} regions. 

Likewise, our identification method for H{\sc ii} regions in narrowband \ha\ maps has an impact on our results. If we insufficiently subtract the DIG, we expect to find an artifical increase of {\em \ha\ only} sightlines at small scales. This, in turn, will drive $f_{scale}$ to artificially low values. Though we inspected our images and verified our algorithm works to first order, adopting a single simple algorithm to analyze galaxy centers, spiral arms, and low density flocculent galaxies will likely yield significant biases. Compounding this situation, the \ha\ line emission is significantly affected by extinction, and both flux-based approaches and the interpretation of thresholds depend on local, robust extinction corrections.

More sophisticated morphological separation methods \citep[e.g., newer versions of H{\sc iiphot};][]{thilker00} offer some prospect to improve the situation. The main hope for improvement here comes from the use of optical integral field spectroscopy (IFS). IFS allows for simultaneous correction for internal extinction and improved tools for distinguishing the \ha\ emission from DIG and shocks or X-ray Dominated Regions (XDRs) associated with
AGN. VLT/MUSE IFS data are currently being obtained for $\sim 19$~PHANGS targets, which have the prospect to then act as a training set for improved morphological and multi-wavelength methods for {\sc Hii} region identification and extinction corrections. Initial tests on the MUSE data for four galaxies yield H{\sc ii} region masks that are more restricted than those constructed in \S\,\ref{subsec:DIG}. This suggests that our {\sc Hii} region identification scheme may bias us to slightly low $f_{scale}$ and thus shorter CO-emitting region lifetimes.

Despite these caveats, we emphasize that most of our analysis focuses on the presence or absence of CO or \ha\ emission, not their exact translation into physical quantities. We made this choice exactly to minimize the bias due to uncertain conversions between CO and H$_2$ or \ha\ and a local star formation rate.


\section{Summary and conclusions}
\label{sec:summary}

We compare high angular resolution observations of CO
line emission to narrowband \ha\ imaging in 
eight nearby star-forming galaxies. We use PHANGS-ALMA CO\,(2-1) imaging for seven
targets and a PdBI CO\,(1-0) map for one target. The \ha\ data come
from a mixture of new and literature sources. Together, the data allow
us to compare the distributions of molecular gas and star formation at
a common spatial scale of 140\,pc. This represents roughly an order of
magnitude increase in sample size compared to previous high resolution comparisons of CO and \ha\ emission, which have mostly compared highly resolved CO and \ha\ in individual Local Group
galaxies.

We use these data to study the spatial relationship between the cold
molecular gas reservoir and \hii\ regions, which serve as signposts of
recent high-mass star formation. We adopt a simple, reproducible
methodology that quantifies overlap between the tracers, in both area
and flux, as a function of scale. We classify each line of sight at
each scale according to its contents: it either contains \textit{only
  CO}, \textit{only \ha }, or \textit{overlapping} CO and
\ha\ emission. We measure the fraction of area in each category as a
function of the resolution of the data. Our main findings are:

\begin{enumerate}

\item {\em CO only} emission (corresponding to $\rm H_2$ surface densities of $\rm \gtrsim 12.6\,M_{\odot}pc^{-2}$) without associated \ha\ emission (above SFR surface densities of 0.0014 and 0.0036\,\msol\,$\rm yr^{-1}\,kpc^{-2}$ depending on the galaxy target) is very common in our sample at $140$~pc resolution. Our targets show a median of $\sim 40\%$ of detected sightlines in this
category, though this numerical value is specific to our adopted thresholds for CO and \ha\ emission. Taken at face value, this implies a large amount of molecular gas without associated recent high-mass star formation. This gas may be in the process of collapsing, may be dynamically unsuited to form stars, or may already be forming low-mass stars.

\item Lines of sight where CO and \ha\ \textit{overlap} represent the
next most common category, also accounting for median $\sim 40\%$ of the
sightlines at 140~pc resolution. That is, \hii\ regions typically coincide with molecular gas at this resolution. Our lowest mass target and our earliest Hubble type system represent notable exceptions to this statement. In those targets, {\sc Hii} region sightlines are the most common type of sightline. We note some concerns related to the impact of completeness, adopted thresholds, and conversion factor on our results for these targets.

\item The balance between {\em CO only}, {\em \ha\ only}, and {\em
  overlapping} lines of sight is a strong function of spatial
  scale. As we degrade the resolution of our maps, the fraction of
  {\em CO only} emission -- tracing apparently quiescent molecular gas --
  decreases, while a steadily increasing fraction of the CO emission becomes associated with {\em overlapping} sightlines. By $1$~kpc resolution, $\sim 75\%$ of all
  sightlines, on average, show overlapping \ha\ and CO emission. At resolutions
  coarser than $\sim 1$\,kpc, the main morphological features of our sample galaxies are
  indistinct. Little information about the relative distributions of
  CO and \ha\ emission or their relation to galactic morphology remains
  at these coarse resolutions. This behavior is consistent with literature results that report a decrease in the scatter of the scaling relations when going from small to large
  scales.

\item At $140$~pc resolution, the spatial distributions of both CO
  emission and \hii\ regions follow galactic structures such as rings,
  spiral arms and bars. CO emission appears to faithfully trace these
  ``backbones'' of the underlying galactic potential. This highlights
  the role of these stellar structures in organizing the molecular gas
  reservoir and star formation. \ha\ emission follows these 
  structures as well, but in a more spotty fashion. Compared to the CO emission, \ha\ emission appears more localized. This patchy appearance of the \hii\ regions
  partially reflects our processing of the \ha\ maps to remove the DIG contribution, and the sensitivity of our \ha\ data.
  
\item Taking the simplest statistical approach, our galaxy-averaged
  measurements of the fractions of sightlines with CO, \ha\ emission or both at $140$~pc resolution imply that, on average, molecular
  gas cycles through a non-star-forming state to a star-forming state
  and then is dispersed 
  on a median timescale that is $1.4\times$ longer than the visibility time for
  \ha\ emission (i.e., $\sim 10{-}15$~Myr). Considering that the physical resolution of our data does not yet attain the characteristic size of molecular clouds and \hii\ regions in the Milky Way, we expect that this timescale represents a lower limit.

\item \ha\ emission appears offset from CO in strong spiral arms, in the sense that the \ha\ emission tends to
  lie along the convex side of the spiral arm compared to the CO
  emission. \ha\ emission is 
  weak or absent along some stellar bars, while sightlines along the (often linear) gas lanes
  along strong bars frequently show {\em CO only} emission. Both these results reflect the influence of dynamical features on the molecular gas distribution and star formation activity in our galaxies. 
  
\item We find some trend between the {\em overlap} fraction
  (computed at 140\,pc resolution) and the fraction of recovered total
  CO flux in the maps. Further we see some evidence for trends relating  stellar mass, SFR, and sSFR to the overall balance between {\em CO only}, {\em \ha\ only}, and {\em overlap} fractions. The sense of the trends is that massive, high-SFR, high-sSFR galaxies tend to show high {\em CO only} fractions. Confirming these
  trends will require analysis of a significantly larger and more diverse sample.

\end{enumerate}

The methodology presented in this paper can be easily applied to much larger
samples, including the full PHANGS-ALMA and paired \ha\ survey now
underway. This larger sample should allow us to identify potential
trends with global galaxy parameters more robustly. 
In particular, more sensitive maps of dwarf spirals and the outer parts of galaxies should help illuminate the nature of systems like NGC\,5068 and NGC\,3351 which showed distinctly different trends from the remaining sample. Other follow-up studies (utilizing a larger sample) include the detailed investigation of how
physical properties of the CO-traced molecular gas vary between star-forming and
non-star-forming regions, and how galactic dynamics drives the suppression of star formation.

A straightforward future extension of the methodology is the implementation of a more generalized form utilizing the information from the cross-distribution of the CO and \ha\ emission (via calculating cross cumulative distribution functions, xCDFs) without the need to establish flux thresholds, and compare this information from xCDFs with those from traditional CDFs. This kind of analysis offers a robust, rich way to characterize the joint distributions of CO
and \ha\ emission in galaxies, yielding measurements ideally suited for
comparison with simulations. As well as trends with galactocentric radius,
comparing the statistics of both traditional and 'cross' distributions among different galactic environments should  provide further insight on how local conditions influence star formation and feedback.


\acknowledgments
This work was carried out as part of the PHANGS collaboration. 
We thank the anonymous referee for constructive feedback that helped improve the manuscript.
We thank B. Venemans for help with the absolute astrometric calibration of the \ha\ maps used for analysis. 
ES, CF, and TS  acknowledge funding from the European Research Council (ERC) under the European Union’s Horizon 2020 research and innovation programme (grant agreement No. 694343).
AH acknowledges support by the Programme National Cosmology et Galaxies (PNCG) of CNRS/INSU with INP and IN2P3, co-funded by CEA and CNES, and by the Programme National “Physique et Chimie du Milieu Interstellaire” (PCMI) of CNRS/INSU with INC/INP co-funded by CEA and CNES.
The work of AKL is partially supported by the National Science Foundation under Grants No.~1615105, 1615109,and 1653300. AKL also acknowledges partial support from NASA ADAP grants NNX16AF48G and NNX17AF39G.
BG gratefully acknowledges the support of the Australian Research Council as the recipient of a Future Fellowship (FT140101202).
KK gratefully acknowledges support from grant KR 4598/1-2 from the DFG Priority Program 1573.  
FB acknowledges funding from the European Research Council (ERC) under the European Union’s Horizon 2020 research and innovation programme (grant agreement No. 726384).
SG acknowledges support from the Deutsche Forschungsgemeinschaft via the Collaborative Research Centre (SFB 881) ``The Milky Way System'' (subprojects B1, B2, and B8) and from the Heidelberg cluster of excellence EXC 2181 ``STRUCTURES: A unifying approach to emergent phenomena in the physical world, mathematics, and complex data'' funded by the German Excellence Strategy.
JMDK and MC acknowledge funding from the German Research Foundation (DFG) in the form of an Emmy Noether Research Group (grant number KR4801/1-1), and the DFG Sachbeihilfe (grant number KR4801/2-1). JMDK acknowledges funding from the European Research Council (ERC) under the European Union’s Horizon 2020 research and innovation programme via the ERC Starting Grant MUSTANG (grant agreement number 714907).
JP acknowledges support from the Programme National “Physique et Chimie du Milieu Interstellaire” (PCMI) of CNRS/INSU with INC/INP co-funded by CEA and CNES.
ER acknowledges the support of the Natural Sciences and Engineering Research Council of Canada (NSERC), funding reference number RGPIN-2017-03987.
The work of JS amd DU is partially supported by the National Science Foundation under Grants No. 1615105, 1615109, and 1653300.
This paper makes use of the following ALMA data: ADS/JAO.ALMA \#2012.1.00650.S, ADS/JAO.ALMA \#2015.1.00925.S, ADS/JAO.ALMA \#2015.1.00956.S. ALMA is a partnership of ESO (representing its member states), NSF (USA) and NINS (Japan), together with NRC (Canada), NSC and ASIAA (Taiwan), and KASI (Republic of Korea), in cooperation with the Republic of Chile. The Joint ALMA Observatory is operated by ESO, AUI/NRAO and NAOJ. The National Radio Astronomy Observatory is a facility of the National Science Foundation operated under cooperative agreement by Associated Universities, Inc. 
This paper makes use of the PdBI Arcsecond Whirlpool Survey \citep{schinnerer13,pety13}. The IRAM 30m telescope and PdBI are run by IRAM, which is supported by INSU/CNRS (France), MPG (Germany) and IGN (Spain).
This work has made use of data from the European Space Agency (ESA) mission
{\it Gaia} (\url{https://www.cosmos.esa.int/gaia}), processed by the {\it Gaia}
Data Processing and Analysis Consortium (DPAC,
\url{https://www.cosmos.esa.int/web/gaia/dpac/consortium}). Funding for the DPAC
has been provided by national institutions, in particular the institutions
participating in the {\it Gaia} Multilateral Agreement.

\facilities{ALMA, IRAM (PdBI and 30m), MPG\,2.2m (WFI)}


\appendix


\section{Excluding the central regions of galaxies}
\label{appendix:piechart_center}

\begin{figure*}[tbh]
\includegraphics[width=\textwidth,angle=0]{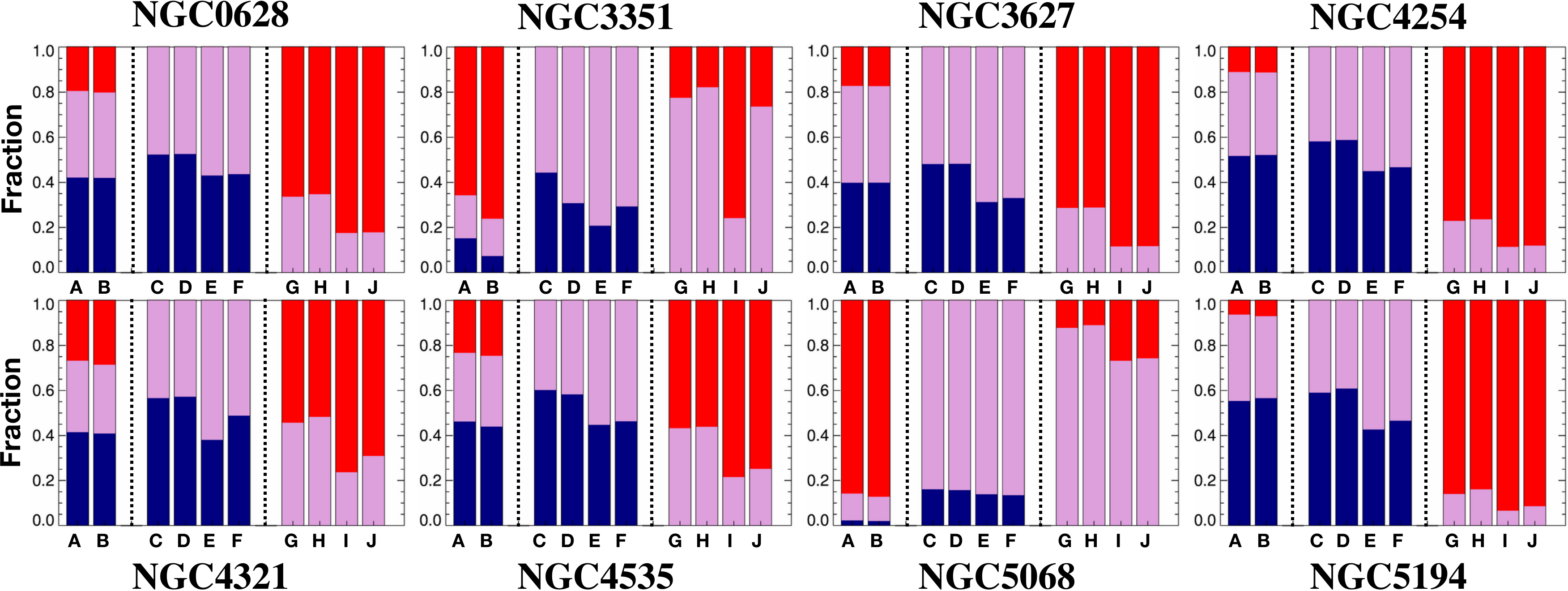}\hfill
\caption{Bar graphs summarizing the impact of different assumptions (sightlines vs. flux; full FoV vs. disk only) for
the different fractions of pixels in our sample galaxies. The color coding is dark blue --
{\em CO only}, red -- {\em \ha\ only}, lavender -- {\em overlap}. 
The bar graphs for each individual galaxy show the following (labels as indicated on the x-axis):
relative distribution of {\em CO only}, {\em \ha\ only} and {\em overlap} sightlines for our FoV ({\bf A}, analoguous to Fig.\,\ref{fig:piechart_gallery_p1})
and disk only ({\bf B});
relative distribution of sightlines tracing CO emission for our FoV ({\bf C}) and disk only ({\bf D}) vs. distribution of CO flux ({\bf E} -- our FoV, {\bf F} -- disk only);
and 
relative distribution of sightlines tracing \ha\ emission for our FoV ({\bf G}) and disk only ({\bf H}) vs. distribution of \ha\ flux ({\bf I} -- our FoV, {\bf J} -- disk only).
\label{fig:appendix_summary}}
\end{figure*}

About half the galaxies in our sample show a non-neglible difference
in the fractions of quiescent and star-forming molecular gas when
these fractions are defined by sightline or by flux (see
\S\,\ref{subsec:2d-maps_140pc}). To test whether CO-bright galaxy
centers are responsible for this difference, we repeat our analysis
presented in that section, excluding the central region from our
measurements. For simplicity, we adopt the central region definition
used by \cite{sun18}, i.e. we define the center as the region
within 1\,kpc of the galaxy reference position (see
Table~\ref{tab:sample}). 
We follow \cite{sun18} who adopted a larger radius of 1.5\,kpc
so that the visually distinct inner disk is entirely inside the central
region
of NGC\,3351. The resulting resulting fractions are
listed in Tab.\,\ref{tab:piechart_disk} and
Tab.\,\ref{tab:byflux_bynum_disk} and compared in Fig.\,\ref{fig:appendix_summary}. As can be seen from the
tables and figure, the centers can explain some of the differences between
fractions measured by sightline or by flux for some galaxies, but not
in all cases.

\begin{deluxetable}{c|ccc}
\centering \tablecaption{Line of Sight Fractions between Tracers at
  140\,pc Scale for Disks Only.\label{tab:piechart_disk}}
\tablehead{
  \colhead{Name} &
  \colhead{CO only} &
  \colhead{\ha\ only} &
  \colhead{overlap} \\
  \colhead{} &
  \colhead{(\%)} &
  \colhead{(\%)} &
  \colhead{(\%)} }
\startdata
NGC0628 & 42 & 20 & 38 \\
NGC3351 & 7 & 76 & 17 \\
NGC3627 & 40 & 17 & 43 \\
NGC4254 & 52 & 11 & 37 \\
NGC4321 & 41 & 29 & 31 \\
NGC4535 & 44 & 25 & 31 \\
NGC5068 & 2 & 87 & 11 \\
NGC5194 & 57 & 7 & 36 \\
\hline
Median & 42 & 25 & 36\\
Mean & 36 & 34 & 30 \\
\enddata
\tablecomments{Fraction of sightlines with CO
  emission only, \ha\ emission only, and both CO and \ha\ emission
  {\em for disks only}, i.e. excluding emission arising inside the
  central region. This table is analogous to Tab.\,\ref{tab:piechart}
  in the main text. Note that the medians are not re-normalized, thus the sum will not add up to 100\%.}
\end{deluxetable}

\begin{deluxetable*}{c|cc|cc|cc|cc}
\centering
\tablecaption{Line of Sight and Flux Fractions per Tracer at 140\,pc Scale.\label{tab:byflux_bynum_disk}}
\tablehead{
\colhead{} &
\multicolumn{4}{|c|}{By Number}  & 
\multicolumn{4}{c}{By Flux} 
\\
\colhead{} &
\multicolumn{2}{|c|}{CO Sightlines}  & 
\multicolumn{2}{c|}{\ha\ Sightlines} &
\multicolumn{2}{c|}{CO Flux}  & 
\multicolumn{2}{c}{\ha\ Flux} 
\\
\multicolumn{1}{c|}{Name} &
\colhead{CO only} &
\multicolumn{1}{c|}{CO overlap} &
\colhead{\ha\ only} &
\multicolumn{1}{c|}{\ha\ overlap} &
\colhead{CO only} &
\multicolumn{1}{c|}{CO overlap} &
\colhead{\ha\ only} &
\colhead{\ha\ overlap} 
\\
\multicolumn{1}{c|}{} &
\colhead{(\%)} &
\multicolumn{1}{c|}{(\%)} &
\colhead{(\%)} &
\multicolumn{1}{c|}{(\%)} &
\colhead{(\%)} &
\multicolumn{1}{c|}{(\%)} &
\colhead{(\%)} &
\colhead{(\%)}
}
\startdata
NGC\,0628 &   51 &   49 &   36 &   64 &   42  &  58  &  20  &  80 \\
NGC\,3351 &   35 &   65 &   74 &   26 &   33  &  67  &  61  &  39 \\
NGC\,3627 &   38 &   62 &   34 &   66 &   23  &  77  &  14  &  86 \\
NGC\,4254 &   58 &   42 &   18 &   82 &   43  &  57  &  10  &  90 \\
NGC\,4321 &   65 &   35 &   40 &   60 &   54  &  46  &  25  &  75 \\
NGC\,4535 &   63 &   37 &   40 &   60 &   50  &  50  &  24  &  76 \\
NGC\,5068 &   20 &   80 &   88 &   12 &   17  &  83  &  68  &  32 \\
NGC\,5194 &   60 &   40 &   17 &   83 &   46  &  54  &  9  &  91 \\
\hline 
Median  &  55  &  45  &  36  &  64  &  42  &  58  &  20  &  80  \\
Mean  &  49  &  51  &  41  &  59  &  39  &  61  &  27  &  73  \\
\enddata
\tablecomments{Fraction of sightlines with {\it CO only} emission,
  {\it \ha\ only} emission, and both CO and \ha\ {\it overlapping}
  emission {\it for disks only}, i.e. excluding emission arising from
  within the central region. This table is analogous to
  Tab.\,\ref{tab:byflux_bynum} in the main text. 
  Note that the medians are not re-normalized, thus the sum will not add up to 100\%.}
\end{deluxetable*}

\end{document}